\documentclass[%
 reprint,
 superscriptaddress,
 amsmath,amssymb,
 aps,
 prl,
longbibliography
]{revtex4-1}

\usepackage{graphicx}
\usepackage{dcolumn}
\usepackage{xcolor}
\usepackage{bm}

\usepackage[colorlinks=true,linkcolor=blue, citecolor=blue]{hyperref}%

\begin{document}

\title{Triplet superconductivity by the orbital Rashba effect at surfaces of elemental superconductors}

\author{Tom G. Saunderson}
\email{t.saunderson@fz-juelich.de}
\affiliation{Institute of Physics, Johannes Gutenberg-University Mainz, Staudingerweg 7, 55128 Mainz, Germany}
\affiliation{Peter Gr\"unberg Institut and Institute for Advanced Simulation, Forschungszentrum J\"ulich and JARA, 52425 J\"ulich, Germany}
\affiliation{Department of Physics, University of South Florida, Tampa, Florida 33620, USA}
\author{Martin Gradhand}
\affiliation{Institute of Physics, Johannes Gutenberg-University Mainz, Staudingerweg 7, 55128 Mainz, Germany}
\author{Dongwook Go}%
\affiliation{Institute of Physics, Johannes Gutenberg-University Mainz, Staudingerweg 7, 55128 Mainz, Germany}
\affiliation{Peter Gr\"unberg Institut and Institute for Advanced Simulation, Forschungszentrum J\"ulich and JARA, 52425 J\"ulich, Germany} 
\author{James F. Annett}
\affiliation{H. H. Wills Physics Laboratory, University of Bristol, Tyndall Avenue, Bristol BS8-1TL, United Kingdom}
\author{Maria Teresa Mercaldo}
\affiliation{Dipartimento di Fisica “E. R. Caianiello,” Universit\`a di Salerno, IT-84084 Fisciano (SA), Italy} 
\author{Mario Cuoco}
\affiliation{Dipartimento di Fisica “E. R. Caianiello,” Universit\`a di Salerno, IT-84084 Fisciano (SA), Italy}
\affiliation{SPIN-CNR, IT-84084 Fisciano (SA), Italy}
\author{Mathias Kl\"aui}
\affiliation{Institute of Physics, Johannes Gutenberg-University Mainz, Staudingerweg 7, 55128 Mainz, Germany}
\affiliation{Centre for Quantum Spintronics, Department of Physics, Norwegian University of Science and Technology, 7491 Trondheim, Norway}
\author{Jacob Gayles}%
\affiliation{Department of Physics, University of South Florida, Tampa, Florida 33620, USA}
\author{Yuriy Mokrousov}%
\affiliation{Institute of Physics, Johannes Gutenberg-University Mainz, Staudingerweg 7, 55128 Mainz, Germany}
\affiliation{Peter Gr\"unberg Institut and Institute for Advanced Simulation, Forschungszentrum J\"ulich and JARA, 52425 J\"ulich, Germany}

\date{\today}

\begin{abstract}

It is often assumed that in a superconductor without spin-triplet pairing, the formation of unconventional spin-triplet densities requires the spin-orbit interaction in combination with either broken inversion symmetry or broken time-reversal symmetry. Here, we show from first principles the existence of supercurrent-driven spin triplet densities on the surface of a variety of simple superconducting materials \textcolor{black}{independent} of spin-orbit coupling. We are able to attribute this phenomenon to the superconducting non-relativistic orbital Rashba Edelstein effect. Furthermore, we find that the spin-orbit induced spin moment is one order of magnitude smaller than the orbital moment, and has a vanishing effect on the total magnitude of the induced triplet density. Our findings imply the existence of a route to generate spin-currents without the use of heavy metals. Additionally, as an orbital moment can couple directly to a magnetic field, it shows that orbital physics is the dominant term that drives the superconducting diode effect. 





\end{abstract}

\maketitle

\begin{figure*}[t!]
\includegraphics*[width=0.9\linewidth,clip]{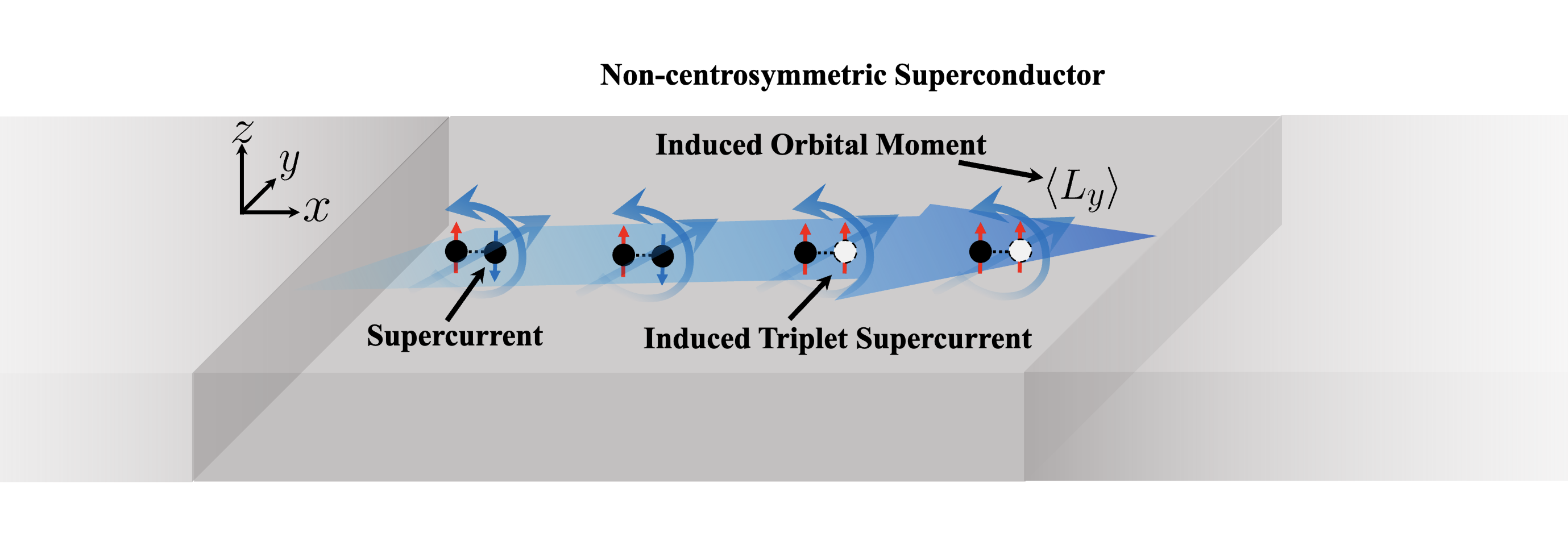}
\caption{A schematic illustration for the superconducting orbital Rashba Edelstein effect on a non-centrosymmetric material. Similarly to the normal state, in the superconducting state there is a super-current induced orbital moment that is driven by symmetry broken orbital Rashba textures on the Fermi surface when passing a current in the $x$-direction. With this setup, the induced orbital moment will be in the y-direction. }
\label{fig:OrbitalRashbaEdelsteinSchematic}
\end{figure*}

While the search for the world's first room temperature superconductor is underway \cite{Lilia2022}, it is clear this phase has the potential to revolutionise more than just the energy industry \cite{Sarma2015,Linder2015,Nadeem2023,Wilson2024,Amundsen2024}. Specifically, whilst the demands for computational power are ever increasing, classical computation suffers from heating and energy loss due to the inevitable finite resistivity introduced when using conventional materials to build computational architectures. The superconducting diode effect \cite{Ando2020,Gutfreund2023,Nadeem2023} is one of the flagship examples which exemplifies the potential to replace a conventional device for one that can perform computations without losses.  

Its physics is understood to be fundamentally linked to the relativistic effect of Rashba spin-orbit coupling \cite{Manchon2015a,Bihlmayer2015}. This effect manifests through structures which exhibit broken inversion symmetry. Probing these systems locally in $k$-space, one finds time reversal symmetry is broken giving rise to spin polarized bands. Globally, such effects are hidden until the application of a current which further breaks the symmetry of the Brillouin zone, giving rise to a global spin polarization. The mechanism plays a crucial role for magnetization dynamics through spin-orbit torques \cite{Manchon2009}, which have subsequently led to a significant number of experimental realizations \cite{Manchon2019}. 

Recently, it has been suggested that there is an effect more fundamental to spin-Rashba splitting, which exists in inversion-broken systems irrespective of spin-orbit coupling. The orbital Rashba effect \cite{Park2011,Park2012,Park2013,Go2017,Go2021,Go2021c} relies on the presence of multiple orbital degrees of freedom, a feature present in the majority of material surfaces. It has been shown to lead to a dramatic enhancement of the current-induced orbital magnetization as compared to its spin counterpart in experimental studies of the spin-orbit torques of Cu films \cite{Ding2020a,Ding2022a,Go2021}. 
The key to this enhancement is the so-called orbital Rashba-Edelstein effect mediated by strong orbital character of the states at the surface acquired as a result of broken inversion symmetry. 
Meanwhile, theoretical models have shown that this effect can be further enhanced when combined with superconductivity \cite{Chirolli2021,Ando2024}. {\color{black} Experimental investigations of superconducting orbitronics are limited. However, experiments on superconducting spintronics have shown a dramatic enhancement of the signal when the metal undergoes the superconducting transition \cite{Linder2015,Jeon2020}}. This implies that superconducting orbitronics might lead to even further enhancement. Orbital moments and their connection to superconductivity have been investigated before \cite{Robbins2020} and it is clear that understanding this interaction will be fundamental for unconventional superconductivity. 

In this Letter, we 
show that in realistic superconductors, namely V, Al and Pb, a supercurrent-driven orbital Rashba Edelstein effect \textcolor{black}{of similar magnitude for each material} can be observed. Moreover, we find that the acquired orbital moments are about one order of magnitude larger than the spin moments for these materials, showcasing the importance of the orbital Rashba Edelstein response over its spin counterpart. \textcolor{black}{Even with 1\% spin-orbit strength}, the superconducting orbital Rashba effect induces $\vert\uparrow\uparrow\rangle$ and   $\vert\downarrow\downarrow\rangle$ triplet densities in these materials. In this Letter we provide the first example of induced spin triplet superconductivity from an orbital response, implying a method to circumnavigate materials with large spin-orbit coupling to generate spin currents. \textcolor{black}{It implies a fundamentally distinct channel that runs parallel to the spin that allows for unconventional superconductivity}. Finally, as an orbital moment can be coupled directly to a magnetic field, we suggest that the orbital moment is the principle component that drives the superconducting diode effect~\cite{Ando2020,Gutfreund2023,Nadeem2023}. 

{\color{black} We employ a first principles Green's function-based density functional theory (DFT) method \cite{Saunderson2020,Saunderson2020b,Saunderson2022,Wu2023} to determine the magnitude of the induced orbital and spin moments of realistic materials. } In order to incorporate the effects of superconductivity, magnetism and spin-orbit coupling we implemented the Dirac Bogoliubov-de Gennes equation \cite{Csire2018} 
\begin{equation}
H_\mathrm{DBdG}(\mathbf{r}) = \left(\begin{matrix}
H_\mathrm{D}(\mathbf{r}) & \Delta_\mathrm{eff}(\mathbf{r}) \\
\Delta^*_\mathrm{eff}(\mathbf{r}) & -H^*_\mathrm{D}(\mathbf{r})
\end{matrix}\right),
\label{eqn:DBdG}
\end{equation}
where $H_\mathrm{D}(\mathbf{r})$ is the $4\times4$ Dirac Hamiltonian and $\Delta_\mathrm{eff}(\mathbf{r})$ denotes the $4\times4$ effective pairing potential of the superconductor which in this instance is considered spin singlet. Despite this, we calculate the three triplet components [$n^{\uparrow\uparrow}(\epsilon)$, $n^{\downarrow\downarrow}(\epsilon)$ and $n^{T_z}(\epsilon)=n^{\uparrow\downarrow}(\epsilon)+n^{\downarrow\uparrow}(\epsilon)$] of the density alongside the singlet component [$n^{s}(\epsilon)=n^{\uparrow\downarrow}(\epsilon)-n^{\downarrow\uparrow}(\epsilon)$] as they are not trivially zero when combining singlet superconductivity with magnetism, spin-orbit coupling and broken inversion symmetry \cite{Sato2017}. This first principles treatment of superconductivity has been used to model unconventional pairing \cite{Csire2018b,Csire2020a,Ghosh2020b}, Rashba superconductivity \cite{Rußmann2022b,Rußmann2023}, Josephson Junctions \cite{Yamazaki2024}, Iron-based superconductivity \cite{Reho2024} and topological superconductivity \cite{Nyari2023,Laszloffy2023} induced from impurities \cite{Saunderson2020b,Saunderson2022,Wu2023} and superconductor/topological insulator interfaces \cite{Rußmann2022a,Reho2024a,Park2020a}. Further details of the first principles treatment of superconductivity can be found in the supplementary information~\cite{Supplementary}. 

\textcolor{black}{The induced supercurrent can be modelled via a modulation of the phase of the superconducting order paramter, such as in a Josephson Junction \cite{Yamazaki2024}. In this instance the symmetry breaking effect of a supercurrent can be modelled as a linear phase gradient \cite{Chirolli2021}, where the setup is displayed in the schematic in Fig.~\ref{fig:OrbitalRashbaEdelsteinSchematic}. We model the effect of a supercurrent on a non-centrosymmetric superconductor by modulating the phase of the order parameter along the $x$-direction. The non-centrosymmetric superconductors chosen for this work are the surfaces of elemental superconductors. With the application of a current along $x$, the induced orbital polarization will be along $y$.}


We assess the orbital Rashba effect of three elemental superconductors, Al, V, and Pb, each with increasing spin-orbit coupling strength, using an ultra-thin two-atom bilayer of the (100) surface of bcc V and the (110) surface of fcc Al and Pb. Whilst this choice of crystal does not necessarily relate to realistic device architecture, this Letter serves as a proof of principle and therefore we choose materials which are commonly used and easy to grow whilst trying to engineer structures that maximize the superconducting orbital Rashba effect. Further information of the computational methodology can be found in the supplementary information \cite{Supplementary}.

\begin{figure}[t]
\includegraphics[width=1.0\linewidth,clip]{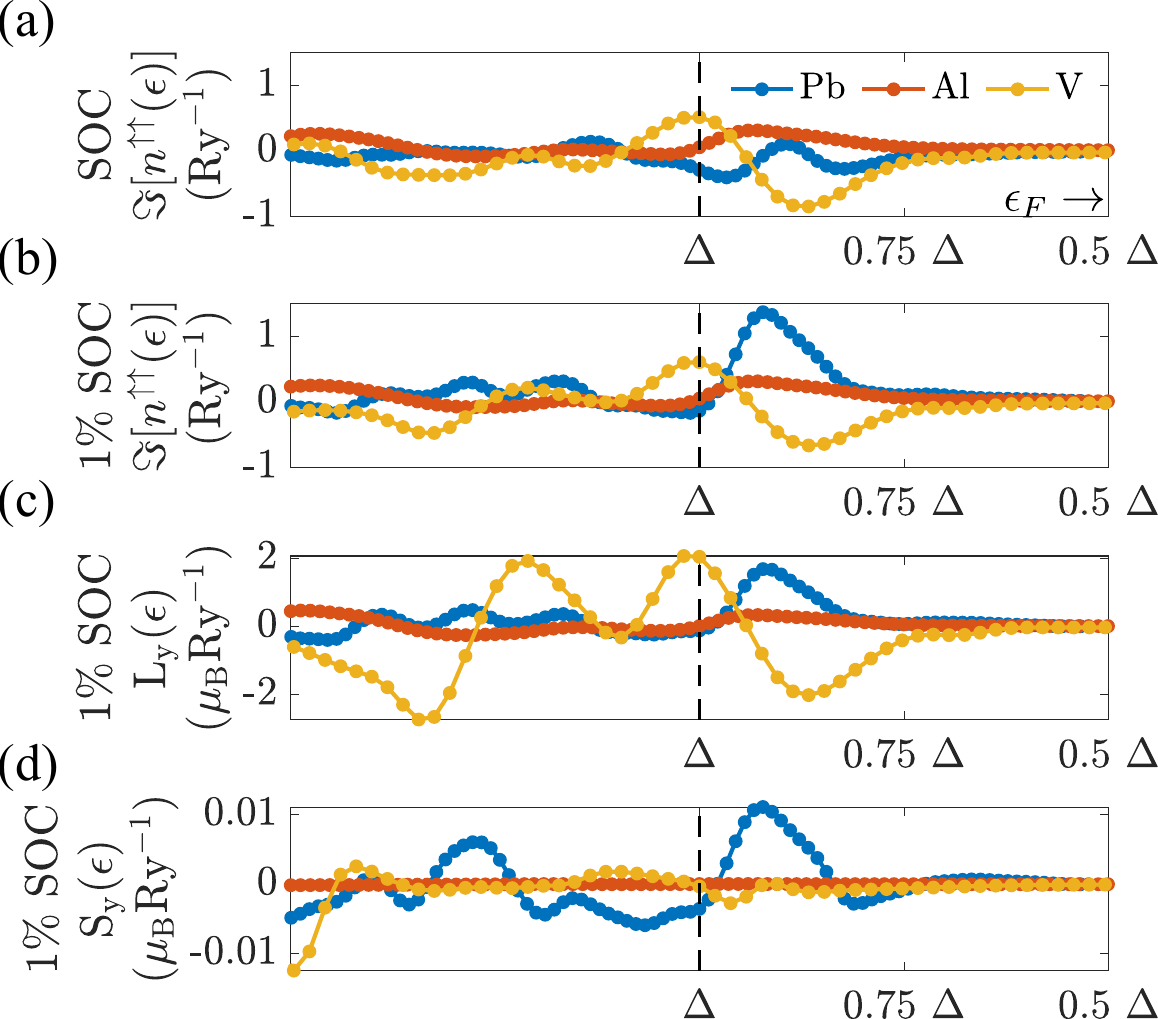}
\caption{Current-induced imaginary part of $n^{\uparrow\uparrow}(\epsilon)$ triplet (a-b) and orbital (c) density around the superconducting gap $\Delta$ for Pb (blue), Al (red) and V (yellow). 
(a) The  current-induced $n^{\uparrow\uparrow}(\epsilon)$ triplet density around the superconducting gap $\Delta$ for Pb (blue), Al (red) and V (yellow) with spin-orbit coupling.  
(b) Same as in (a) with only 1\% spin-orbit coupling strength.
\textcolor{black}{(c) The $y$-component of the current-induced orbital ($L_y$) moment around the superconducting gap for Pb (blue), Al (red) and V (yellow) with 1\% spin-orbit coupling.
(d) The $y$-component of the current-induced spin ($S_y$) moment around the superconducting gap for Pb (blue), Al (red) and V (yellow) with 1\% spin-orbit coupling.}
} 
\label{fig:PlotDOS_NoSOCvSOCvOrbitalNoSOC}
\end{figure}

We first compute the magnitude of the $n^{\uparrow\uparrow}(\epsilon)$ triplet densities \textcolor{black}{for the selected materials} in Fig.~\ref{fig:PlotDOS_NoSOCvSOCvOrbitalNoSOC}(a). As expected, triplet densities are induced for each material in Fig.~\ref{fig:PlotDOS_NoSOCvSOCvOrbitalNoSOC}(a) \textcolor{black}{which remarkably seem to be irrespective of spin-orbit strength}. \textcolor{black}{Scaling the spin-orbit parameter to 1\% in Fig.~\ref{fig:PlotDOS_NoSOCvSOCvOrbitalNoSOC}(b) we see that the triplet density is most greatly affected in Pb, and the least affected in Al. This is consistent with the fact that SOC is much stronger in Pb than in Al, but rejects the conventional understanding that SOC will tune the strength of the effect.} To understand the microscopic origin of the spectrum shown in Fig.~\ref{fig:PlotDOS_NoSOCvSOCvOrbitalNoSOC}(b) we present the $y$ component of the orbital $L_y(\epsilon)$ and spin $S_y(\epsilon)$ moments as a function of band filling in Fig.~\ref{fig:PlotDOS_NoSOCvSOCvOrbitalNoSOC}(c) and (d). \textcolor{black}{Firstly, the scale of figure (d) is two orders of magnitude smaller than (c), which is the expected result with such a small spin orbit strength, and implies that its influence on the triplet density is immaterial.} \textcolor{black}{Comparing Fig.~\ref{fig:PlotDOS_NoSOCvSOCvOrbitalNoSOC}(b) with (c) we see a strong correlation implying that the orbital moment plays the crucial role in determining the strength of the induced triplet density in the limit of low spin orbit strength.} \textcolor{black}{This almost implies that there are effectively two channels to that drive the triplet density in noncentrosymmetric superconductors.
}


\begin{figure}[h]
\includegraphics[width=0.8\linewidth,clip]{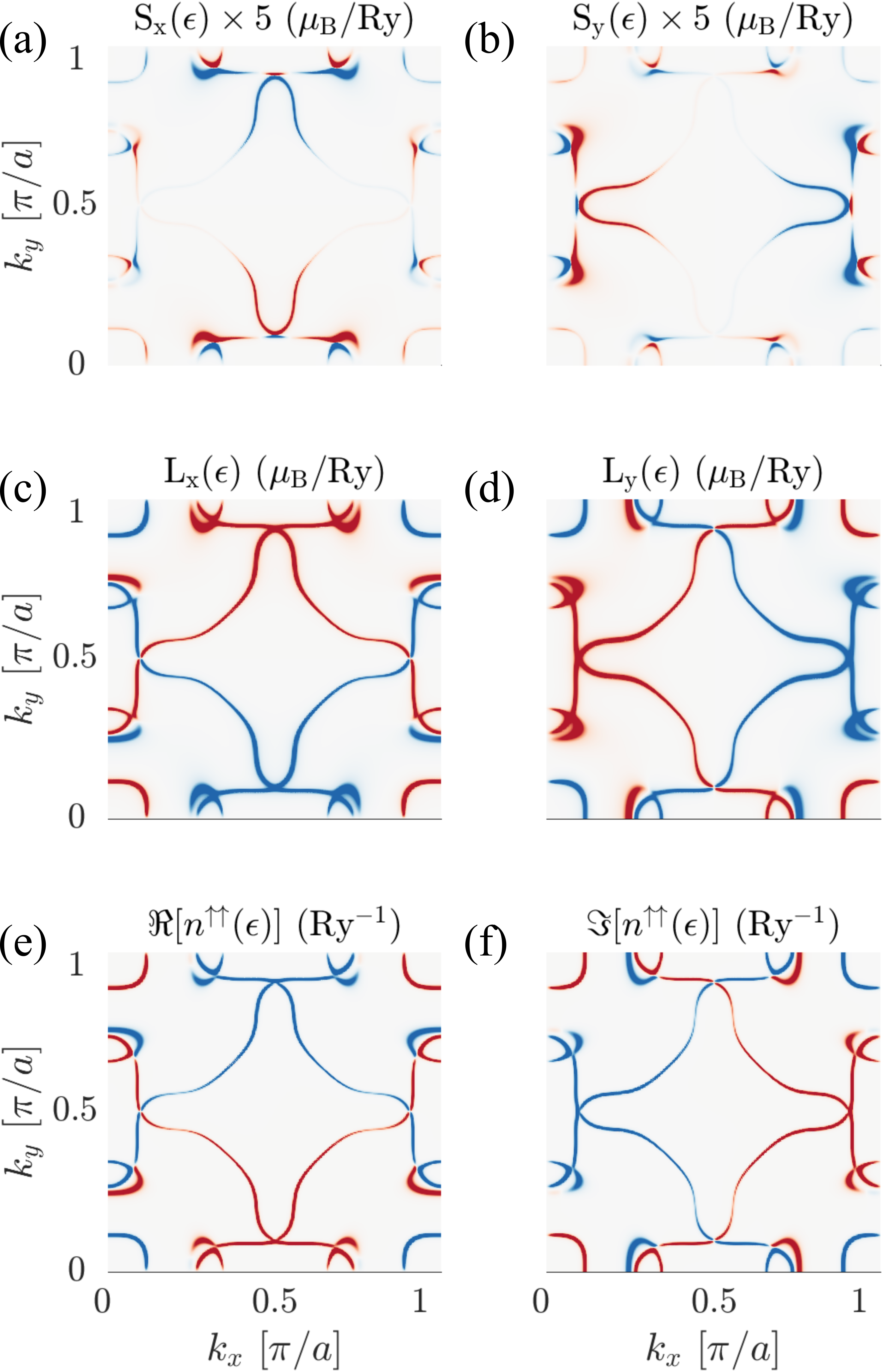}
\caption{The $\mathbf{k}$-space distribution of $\mathbf{S}_{\mathbf{k}}$ (a-b), $\mathbf{L}_{\mathbf{k}}$ (c-d) and $n^{\uparrow\uparrow}_{\mathbf{k}}(\epsilon)$ (e-f) for Vanadium with spin-orbit coupling.  } 
\label{fig:FermiSurface}
\end{figure}

To understand the contributions arising from the orbital and spin Rashba effects in determining the strength of the induced triplet density, we investigate the $\mathbf{k}$-resolved triplet, orbital and spin densities for V at the Fermi surface with spin orbit coupling. Comparing Figs.~\ref{fig:FermiSurface}(a-b) and (c-d) we find the magnitude of the spin component is reduced compared to its orbital counterpart. In fact the magnitude of $S_\sigma(\epsilon)$ is scaled by a factor of 5 in (a) and (b) in comparison to $L_\sigma(\epsilon)$ in (c) and (d). Turning to the triplet density in Fig.~\ref{fig:FermiSurface}(e-f), we find a strong correlation between $L_x(\epsilon)$ and $\Re[n^{\uparrow\uparrow}(\epsilon)]$ [Figs.~\ref{fig:FermiSurface}(c) and (e)], and $L_y(\epsilon)$ and $\Im[n^{\uparrow\uparrow}(\epsilon)]$ [Figs.~\ref{fig:FermiSurface}(d) and (f)], implying a fundamental dependence between the Fermi surface orbital and triplet contributions. In contrast, Figs.~\ref{fig:FermiSurface}(a-b) and (e-f) are not closely related, differing in symmetry, signifying that the orbital Rashba effect is the dominant contribution that drives the triplet density. 

We quantify the induced spin and orbital moment for all materials in Table \ref{tab:OrbAndSpin}. It is clear the induced orbital and spin moments are small, which is in line with the conventional understanding of Rashba Edelstein effects. What is more important are the relative magnitudes of the current-induced spin and orbital moments. The spin-orbit strength decreases from left to right in the table. Looking at the ratio between spin and orbital moment it is clear that the induced spin moment follows the same trend, namely being suppressed as we go from Pb to Al. It is important to note that even with Pb, the material with the highest spin-orbit coupling, the spin moment is not the dominant contribution. This highlights the importance of the orbital Rashba effect over the spin Rashba effect in superconducting diode experiments once again. 

\begin{figure}[t]
\includegraphics[width=0.8\linewidth,clip]{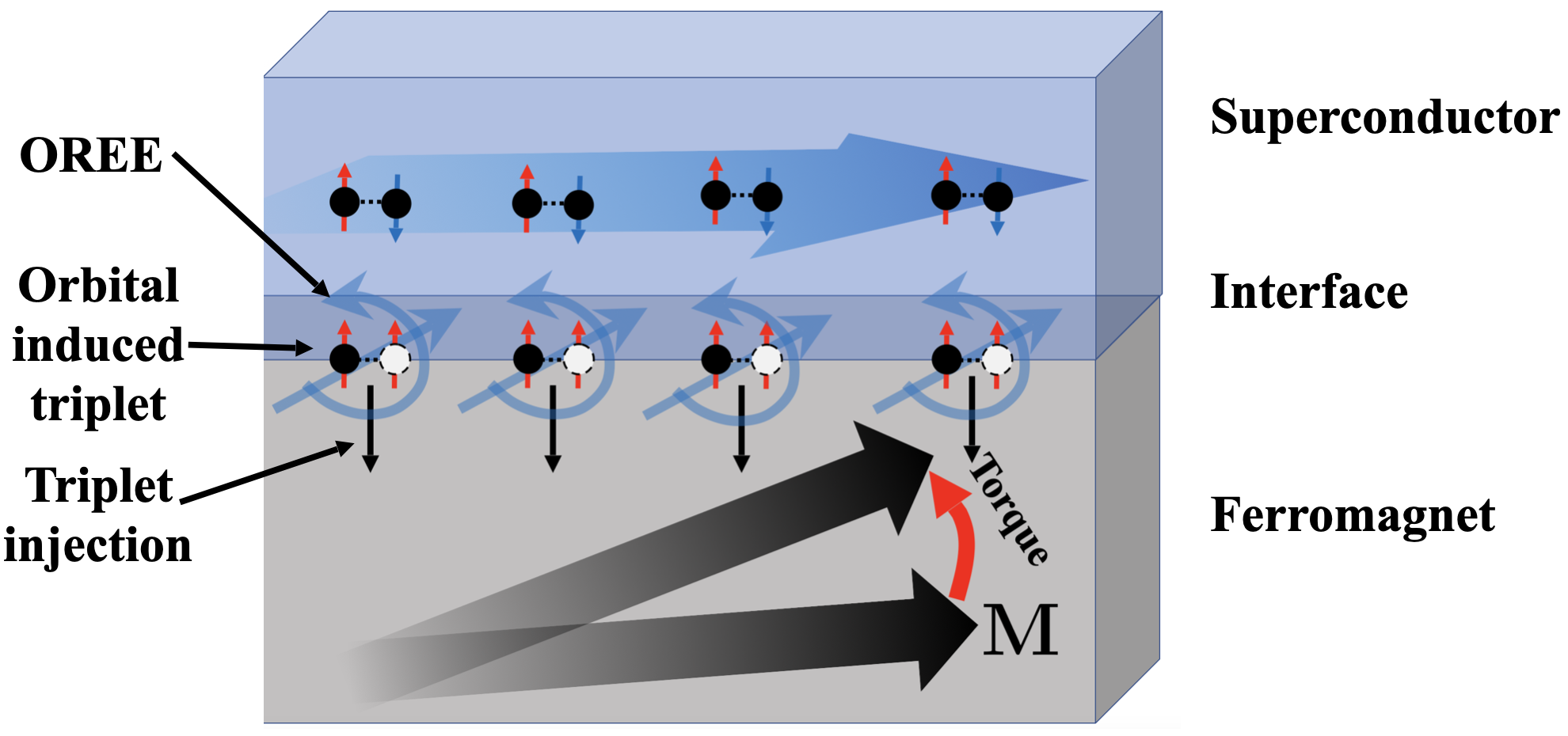}
\caption{A schematic illustration of nonlocal generation of SOTs in FM/SC structures utilizing the superconducting OREE effect at the interface between the SC and FM. The OREE effect from the interface will generate triplet states in the superconductor, which will in turn be injected into the Ferromagnet.  
} 
\label{fig:Experiment_supOREE}
\end{figure}

The first experiment we address are orbital and spin-orbit torques \cite{Manchon2009}. Here, orbital torques have already been investigated in the normal state and have shown significant enhancements compared to conventional spin-orbit torques in CuO/Cu/Pt/FM heterostructures \cite{Ding2020a}. In this experiment, the Pt films provide spin-orbit coupling to convert an orbital current into a spin current. Here, we suggest a new mechanism. In this case, rather than relying on a heavy metal to create a spin-current, we can rely on a superconductor whose response to an injected orbital current will be to generate spin triplet states which will be injected into the ferromagnet. These CuO$_x$/V/FM heterostructures, as illustrated in Fig.~\ref{fig:Experiment_supOREE}, will exhibit vanishing spin-orbit strength and the effect can be tested when sweeping through the superconducting transition temperature of Vanadium.  

\begin{table}
\caption{Current induced spin and orbital moments and their ratio from the ultrathin bilayers of Pb, V and Al.}
\label{tab:OrbAndSpin}
\begin{tabular}{|c||c|c|c|}
 \hline
 \textbf{Moment} ($\mu_B$) & \textbf{Pb} & \textbf{V} & \textbf{Al} \\
 \hline
 \textbf{Spin} & 2.78E-05 & -3.17E-05 & -9.34E-07 \\  
 \hline
 \textbf{Orbital} & 7.31E-05 & -3.24E-04 & 3.42E-05 \\
 \hline
 \textbf{Spin/Orbital} & 38\% & 10\% & -3\% \\
 \hline 
\end{tabular}
\end{table}

The second experiment is the superconducting diode effect \cite{Ando2020,Gutfreund2023,Nadeem2023}. To perform this experiment, one sweeps a current in the $\pm x$ direction of the sample while keeping a constant magnetic field applied in the $y$ direction. \textcolor{black}{The Rashba textures present in the system induce a y-polarized moment as exemplified in this work. The constant applied magnetic field therefore acts as a symmetry breaking force which interacts with the induced moment which will change sign depending on whether the current is swept in the $+x$ or the $-x$ direction.} The result will be nonreciprocal behaviour between the critical current in the $+x$ direction versus the critical current in the $-x$ direction. This effect relies on the superconductor breaking inversion symmetry, giving rise to a global $y$ spin-polarization which breaks the symmetry of a $\pm x$ current sweep. Our findings show that the $y$ polarized orbital moment is significantly larger, even in Pb, than the spin moment. As an orbital moment can couple directly to the magnetic field it has its own diode effect which is distinct, and significantly more prominent, than the spin response. \textcolor{black}{This means that, whilst the current experimental search has been to increase the atomic number in order to enhance the spin-orbit coupling, we suggest that this is not the most important effect. For example in the seminal work by F. Ando \textit{et al} \cite{Ando2020}, the superconducting diode effect was observed in Nb/Ta/V multilayers, whilst we suggest that in fact light elemental superconductors will already be sufficient.} 

To conclude, in this work we have shown that there is a coupling channel which exists between the orbital Rashba Edelstein effect and spin triplet Cooper pairs which circumnavigates spin Rashba spin-orbit coupling. This theoretical investigation was performed by modelling a supercurrent on the surface of the realistic superconductors V, Al and Pb.
We show that while the spin-Rashba effect is present, the orbital Rashba effect is the dominant quantity which induces spin triplet densities even \textcolor{black}{1\%} spin-orbit coupling. Furthermore, the current-induced orbital moment is consistently higher than the current-induced spin moment. Finally, two experimental setups were discussed that could realise this theoretical discovery. Such discoveries open up a new chapter of unconventional superconductivity. One such example would be the question of orbital topology and its interaction with superconductivity would generate new forms of topological superconductivity. Another such example would be the interaction with chiral phonons and crystals whose higher orbital angular momentum gives rise t o superconductivity with higher degrees of freedom. Finally, this discovery provides new ways to incorporate superconductors into hybrid structures with magnetic systems, providing the blueprint for the next generation of computational architectures.  

\begin{acknowledgments}
The authors appreciate fruitful discussions with Prof. Bal\'azs Ujfalussy, Dr. Philipp R\"u{\ss}mann and Dr. Gabor Csire. 
This work was supported by the EIC Pathfinder OPEN grant 101129641 ``OBELIX''. We also  acknowledge the funding by the Deutsche Forschungsgemeinschaft TRR 173/3—268565370 (Projects A01 and A11), TRR 288—422213477 (Project B06). We gratefully acknowledge the J\"ulich Supercomputing Centre for providing computational resources under project jiff40. TGS gratefully acknowledges funding from TopDyn - Dynamics and Topology early career researchers grant, and acknowledges support from the US National Science Foundation (NSF) Grant Number 2201516 under the Accelnet program of Office of International Science and Engineering (OISE). This material is based upon work supported by the Air
Force Office of Scientific Research under award number
FA9550-23-1-0132.  
\end{acknowledgments}

\bibliography{SuperconductingOrbitalRashba_Paper}

\begin{thebibliography}{50}%
\makeatletter
\providecommand \@ifxundefined [1]{%
 \@ifx{#1\undefined}
}%
\providecommand \@ifnum [1]{%
 \ifnum #1\expandafter \@firstoftwo
 \else \expandafter \@secondoftwo
 \fi
}%
\providecommand \@ifx [1]{%
 \ifx #1\expandafter \@firstoftwo
 \else \expandafter \@secondoftwo
 \fi
}%
\providecommand \natexlab [1]{#1}%
\providecommand \enquote  [1]{``#1''}%
\providecommand \bibnamefont  [1]{#1}%
\providecommand \bibfnamefont [1]{#1}%
\providecommand \citenamefont [1]{#1}%
\providecommand \href@noop [0]{\@secondoftwo}%
\providecommand \href [0]{\begingroup \@sanitize@url \@href}%
\providecommand \@href[1]{\@@startlink{#1}\@@href}%
\providecommand \@@href[1]{\endgroup#1\@@endlink}%
\providecommand \@sanitize@url [0]{\catcode `\\12\catcode `\$12\catcode `\&12\catcode `\#12\catcode `\^12\catcode `\_12\catcode `\%12\relax}%
\providecommand \@@startlink[1]{}%
\providecommand \@@endlink[0]{}%
\providecommand \url  [0]{\begingroup\@sanitize@url \@url }%
\providecommand \@url [1]{\endgroup\@href {#1}{\urlprefix }}%
\providecommand \urlprefix  [0]{URL }%
\providecommand \Eprint [0]{\href }%
\providecommand \doibase [0]{http://dx.doi.org/}%
\providecommand \selectlanguage [0]{\@gobble}%
\providecommand \bibinfo  [0]{\@secondoftwo}%
\providecommand \bibfield  [0]{\@secondoftwo}%
\providecommand \translation [1]{[#1]}%
\providecommand \BibitemOpen [0]{}%
\providecommand \bibitemStop [0]{}%
\providecommand \bibitemNoStop [0]{.\EOS\space}%
\providecommand \EOS [0]{\spacefactor3000\relax}%
\providecommand \BibitemShut  [1]{\csname bibitem#1\endcsname}%
\let\auto@bib@innerbib\@empty
\bibitem [{\citenamefont {Lilia}\ \emph {et~al.}(2022)\citenamefont {Lilia}, \citenamefont {Hennig}, \citenamefont {Hirschfeld}, \citenamefont {Profeta}, \citenamefont {Sanna}, \citenamefont {Zurek}, \citenamefont {Pickett}, \citenamefont {Amsler}, \citenamefont {Dias}, \citenamefont {Eremets}, \citenamefont {Heil}, \citenamefont {Hemley}, \citenamefont {Liu}, \citenamefont {Ma}, \citenamefont {Pierleoni}, \citenamefont {Kolmogorov}, \citenamefont {Rybin}, \citenamefont {Novoselov}, \citenamefont {Anisimov}, \citenamefont {Oganov}, \citenamefont {Pickard}, \citenamefont {Bi}, \citenamefont {Arita}, \citenamefont {Errea}, \citenamefont {Pellegrini}, \citenamefont {Requist}, \citenamefont {Gross}, \citenamefont {Margine}, \citenamefont {Xie}, \citenamefont {Quan}, \citenamefont {Hire}, \citenamefont {Fanfarillo}, \citenamefont {Stewart}, \citenamefont {Hamlin}, \citenamefont {Stanev}, \citenamefont {Gonnelli}, \citenamefont {Piatti}, \citenamefont {Romanin}, \citenamefont {Daghero},\ and\ \citenamefont {Valenti}}]{Lilia2022}%
  \BibitemOpen
  \bibfield  {author} {\bibinfo {author} {\bibfnamefont {Boeri}\ \bibnamefont {Lilia}}, \bibinfo {author} {\bibfnamefont {Richard}\ \bibnamefont {Hennig}}, \bibinfo {author} {\bibfnamefont {Peter}\ \bibnamefont {Hirschfeld}}, \bibinfo {author} {\bibfnamefont {Gianni}\ \bibnamefont {Profeta}}, \bibinfo {author} {\bibfnamefont {Antonio}\ \bibnamefont {Sanna}}, \bibinfo {author} {\bibfnamefont {Eva}\ \bibnamefont {Zurek}}, \bibinfo {author} {\bibfnamefont {Warren~E.}\ \bibnamefont {Pickett}}, \bibinfo {author} {\bibfnamefont {Maximilian}\ \bibnamefont {Amsler}}, \bibinfo {author} {\bibfnamefont {Ranga}\ \bibnamefont {Dias}}, \bibinfo {author} {\bibfnamefont {Mikhail~I.}\ \bibnamefont {Eremets}}, \bibinfo {author} {\bibfnamefont {Christoph}\ \bibnamefont {Heil}}, \bibinfo {author} {\bibfnamefont {Russell~J.}\ \bibnamefont {Hemley}}, \bibinfo {author} {\bibfnamefont {Hanyu}\ \bibnamefont {Liu}}, \bibinfo {author} {\bibfnamefont {Yanming}\ \bibnamefont {Ma}}, \bibinfo {author} {\bibfnamefont {Carlo}\ \bibnamefont {Pierleoni}}, \bibinfo {author} {\bibfnamefont {Aleksey~N.}\ \bibnamefont {Kolmogorov}}, \bibinfo {author} {\bibfnamefont {Nikita}\ \bibnamefont {Rybin}}, \bibinfo {author} {\bibfnamefont {Dmitry}\ \bibnamefont {Novoselov}}, \bibinfo {author} {\bibfnamefont {Vladimir}\ \bibnamefont {Anisimov}}, \bibinfo {author} {\bibfnamefont {Artem~R.}\ \bibnamefont {Oganov}}, \bibinfo {author} {\bibfnamefont {Chris~J.}\ \bibnamefont {Pickard}}, \bibinfo {author} {\bibfnamefont {Tiange}\ \bibnamefont {Bi}}, \bibinfo {author} {\bibfnamefont {Ryotaro}\ \bibnamefont {Arita}}, \bibinfo {author} {\bibfnamefont {Ion}\ \bibnamefont {Errea}}, \bibinfo {author} {\bibfnamefont {Camilla}\ \bibnamefont {Pellegrini}}, \bibinfo {author} {\bibfnamefont {Ryan}\ \bibnamefont {Requist}}, \bibinfo {author} {\bibfnamefont {E.~K.U.}\ \bibnamefont {Gross}}, \bibinfo {author} {\bibfnamefont {Elena~Roxana}\ \bibnamefont {Margine}}, \bibinfo {author} {\bibfnamefont {Stephen~R.}\ \bibnamefont {Xie}}, \bibinfo {author} {\bibfnamefont {Yundi}\ \bibnamefont {Quan}}, \bibinfo {author} {\bibfnamefont {Ajinkya}\ \bibnamefont {Hire}}, \bibinfo {author} {\bibfnamefont {Laura}\ \bibnamefont {Fanfarillo}}, \bibinfo {author} {\bibfnamefont {G.~R.}\ \bibnamefont {Stewart}}, \bibinfo {author} {\bibfnamefont {J.~J.}\ \bibnamefont {Hamlin}}, \bibinfo {author} {\bibfnamefont {Valentin}\ \bibnamefont {Stanev}}, \bibinfo {author} {\bibfnamefont {Renato~S.}\ \bibnamefont {Gonnelli}}, \bibinfo {author} {\bibfnamefont {Erik}\ \bibnamefont {Piatti}}, \bibinfo {author} {\bibfnamefont {Davide}\ \bibnamefont {Romanin}}, \bibinfo {author} {\bibfnamefont {Dario}\ \bibnamefont {Daghero}}, \ and\ \bibinfo {author} {\bibfnamefont {Roser}\ \bibnamefont {Valenti}},\ }\bibfield  {title} {\enquote {\bibinfo {title} {{The 2021 room-temperature superconductivity roadmap}},}\ }\href {\doibase 10.1088/1361-648X/AC2864} {\bibfield  {journal} {\bibinfo  {journal} {J. Phys. Condens. Matter}\ }\textbf {\bibinfo {volume} {34}},\ \bibinfo {pages} {183002} (\bibinfo {year} {2022})}\BibitemShut {NoStop}%
\bibitem [{\citenamefont {Sarma}\ \emph {et~al.}(2015)\citenamefont {Sarma}, \citenamefont {Freedman},\ and\ \citenamefont {Nayak}}]{Sarma2015}%
  \BibitemOpen
  \bibfield  {author} {\bibinfo {author} {\bibfnamefont {Sankar~Das}\ \bibnamefont {Sarma}}, \bibinfo {author} {\bibfnamefont {Michael}\ \bibnamefont {Freedman}}, \ and\ \bibinfo {author} {\bibfnamefont {Chetan}\ \bibnamefont {Nayak}},\ }\bibfield  {title} {\enquote {\bibinfo {title} {{Majorana zero modes and topological quantum computation}},}\ }\href {\doibase 10.1038/npjqi.2015.1} {\bibfield  {journal} {\bibinfo  {journal} {npj Quantum Inf.}\ }\textbf {\bibinfo {volume} {1}},\ \bibinfo {pages} {15001} (\bibinfo {year} {2015})},\ \Eprint {http://arxiv.org/abs/1501.02813} {arXiv:1501.02813} \BibitemShut {NoStop}%
\bibitem [{\citenamefont {Linder}\ and\ \citenamefont {Robinson}(2015)}]{Linder2015}%
  \BibitemOpen
  \bibfield  {author} {\bibinfo {author} {\bibfnamefont {J.}~\bibnamefont {Linder}}\ and\ \bibinfo {author} {\bibfnamefont {J.~W.~A.}\ \bibnamefont {Robinson}},\ }\bibfield  {title} {\enquote {\bibinfo {title} {{Superconducting spintronics}},}\ }\href {\doibase 10.1038/nphys3242} {\bibfield  {journal} {\bibinfo  {journal} {Nat. Phys.}\ }\textbf {\bibinfo {volume} {11}},\ \bibinfo {pages} {307--315} (\bibinfo {year} {2015})}\BibitemShut {NoStop}%
\bibitem [{\citenamefont {Nadeem}\ \emph {et~al.}(2023)\citenamefont {Nadeem}, \citenamefont {Fuhrer},\ and\ \citenamefont {Wang}}]{Nadeem2023}%
  \BibitemOpen
  \bibfield  {author} {\bibinfo {author} {\bibfnamefont {Muhammad}\ \bibnamefont {Nadeem}}, \bibinfo {author} {\bibfnamefont {Michael~S.}\ \bibnamefont {Fuhrer}}, \ and\ \bibinfo {author} {\bibfnamefont {Xiaolin}\ \bibnamefont {Wang}},\ }\bibfield  {title} {\enquote {\bibinfo {title} {{The superconducting diode effect}},}\ }\href {\doibase 10.1038/s42254-023-00632-w} {\bibfield  {journal} {\bibinfo  {journal} {Nat. Rev. Phys.}\ }\textbf {\bibinfo {volume} {5}},\ \bibinfo {pages} {558--577} (\bibinfo {year} {2023})}\BibitemShut {NoStop}%
\bibitem [{\citenamefont {Wilson}\ and\ \citenamefont {Ortiz}(2024)}]{Wilson2024}%
  \BibitemOpen
  \bibfield  {author} {\bibinfo {author} {\bibfnamefont {Stephen~D.}\ \bibnamefont {Wilson}}\ and\ \bibinfo {author} {\bibfnamefont {Brenden~R.}\ \bibnamefont {Ortiz}},\ }\bibfield  {title} {\enquote {\bibinfo {title} {{AV3Sb5 kagome superconductors}},}\ }\href {\doibase 10.1038/s41578-024-00677-y} {\bibfield  {journal} {\bibinfo  {journal} {Nat. Rev. Mater.}\ }\textbf {\bibinfo {volume} {9}},\ \bibinfo {pages} {420--432} (\bibinfo {year} {2024})}\BibitemShut {NoStop}%
\bibitem [{\citenamefont {Amundsen}\ \emph {et~al.}(2024)\citenamefont {Amundsen}, \citenamefont {Linder}, \citenamefont {Robinson}, \citenamefont {{\v{Z}}uti{\'{c}}},\ and\ \citenamefont {Banerjee}}]{Amundsen2024}%
  \BibitemOpen
  \bibfield  {author} {\bibinfo {author} {\bibfnamefont {Morten}\ \bibnamefont {Amundsen}}, \bibinfo {author} {\bibfnamefont {Jacob}\ \bibnamefont {Linder}}, \bibinfo {author} {\bibfnamefont {Jason~W.A.}\ \bibnamefont {Robinson}}, \bibinfo {author} {\bibfnamefont {Igor}\ \bibnamefont {{\v{Z}}uti{\'{c}}}}, \ and\ \bibinfo {author} {\bibfnamefont {Niladri}\ \bibnamefont {Banerjee}},\ }\bibfield  {title} {\enquote {\bibinfo {title} {{Colloquium: Spin-orbit effects in superconducting hybrid structures}},}\ }\href {\doibase 10.1103/REVMODPHYS.96.021003/FIGURES/9/THUMBNAIL} {\bibfield  {journal} {\bibinfo  {journal} {Rev. Mod. Phys.}\ }\textbf {\bibinfo {volume} {96}},\ \bibinfo {pages} {021003} (\bibinfo {year} {2024})},\ \Eprint {http://arxiv.org/abs/2210.03549} {arXiv:2210.03549} \BibitemShut {NoStop}%
\bibitem [{\citenamefont {Ando}\ \emph {et~al.}(2020)\citenamefont {Ando}, \citenamefont {Miyasaka}, \citenamefont {Li}, \citenamefont {Ishizuka}, \citenamefont {Arakawa}, \citenamefont {Shiota}, \citenamefont {Moriyama}, \citenamefont {Yanase},\ and\ \citenamefont {Ono}}]{Ando2020}%
  \BibitemOpen
  \bibfield  {author} {\bibinfo {author} {\bibfnamefont {Fuyuki}\ \bibnamefont {Ando}}, \bibinfo {author} {\bibfnamefont {Yuta}\ \bibnamefont {Miyasaka}}, \bibinfo {author} {\bibfnamefont {Tian}\ \bibnamefont {Li}}, \bibinfo {author} {\bibfnamefont {Jun}\ \bibnamefont {Ishizuka}}, \bibinfo {author} {\bibfnamefont {Tomonori}\ \bibnamefont {Arakawa}}, \bibinfo {author} {\bibfnamefont {Yoichi}\ \bibnamefont {Shiota}}, \bibinfo {author} {\bibfnamefont {Takahiro}\ \bibnamefont {Moriyama}}, \bibinfo {author} {\bibfnamefont {Youichi}\ \bibnamefont {Yanase}}, \ and\ \bibinfo {author} {\bibfnamefont {Teruo}\ \bibnamefont {Ono}},\ }\bibfield  {title} {\enquote {\bibinfo {title} {{Observation of superconducting diode effect}},}\ }\href {\doibase 10.1038/s41586-020-2590-4} {\bibfield  {journal} {\bibinfo  {journal} {Nature}\ }\textbf {\bibinfo {volume} {584}},\ \bibinfo {pages} {373--376} (\bibinfo {year} {2020})}\BibitemShut {NoStop}%
\bibitem [{\citenamefont {Gutfreund}\ \emph {et~al.}(2023)\citenamefont {Gutfreund}, \citenamefont {Matsuki}, \citenamefont {Plastovets}, \citenamefont {Noah}, \citenamefont {Gorzawski}, \citenamefont {Fridman}, \citenamefont {Yang}, \citenamefont {Buzdin}, \citenamefont {Millo}, \citenamefont {Robinson},\ and\ \citenamefont {Anahory}}]{Gutfreund2023}%
  \BibitemOpen
  \bibfield  {author} {\bibinfo {author} {\bibfnamefont {Alon}\ \bibnamefont {Gutfreund}}, \bibinfo {author} {\bibfnamefont {Hisakazu}\ \bibnamefont {Matsuki}}, \bibinfo {author} {\bibfnamefont {Vadim}\ \bibnamefont {Plastovets}}, \bibinfo {author} {\bibfnamefont {Avia}\ \bibnamefont {Noah}}, \bibinfo {author} {\bibfnamefont {Laura}\ \bibnamefont {Gorzawski}}, \bibinfo {author} {\bibfnamefont {Nofar}\ \bibnamefont {Fridman}}, \bibinfo {author} {\bibfnamefont {Guang}\ \bibnamefont {Yang}}, \bibinfo {author} {\bibfnamefont {Alexander}\ \bibnamefont {Buzdin}}, \bibinfo {author} {\bibfnamefont {Oded}\ \bibnamefont {Millo}}, \bibinfo {author} {\bibfnamefont {Jason~W.A.}\ \bibnamefont {Robinson}}, \ and\ \bibinfo {author} {\bibfnamefont {Yonathan}\ \bibnamefont {Anahory}},\ }\bibfield  {title} {\enquote {\bibinfo {title} {{Direct observation of a superconducting vortex diode}},}\ }\href {\doibase 10.1038/s41467-023-37294-2} {\bibfield  {journal} {\bibinfo  {journal} {Nat. Commun.}\ }\textbf {\bibinfo {volume} {14}},\ \bibinfo {pages} {1630} (\bibinfo {year} {2023})},\ \Eprint {http://arxiv.org/abs/2301.07121} {arXiv:2301.07121} \BibitemShut {NoStop}%
\bibitem [{\citenamefont {Manchon}\ \emph {et~al.}(2015)\citenamefont {Manchon}, \citenamefont {Koo}, \citenamefont {Nitta}, \citenamefont {Frolov},\ and\ \citenamefont {Duine}}]{Manchon2015a}%
  \BibitemOpen
  \bibfield  {author} {\bibinfo {author} {\bibfnamefont {A.}~\bibnamefont {Manchon}}, \bibinfo {author} {\bibfnamefont {H.~C.}\ \bibnamefont {Koo}}, \bibinfo {author} {\bibfnamefont {J.}~\bibnamefont {Nitta}}, \bibinfo {author} {\bibfnamefont {S.~M.}\ \bibnamefont {Frolov}}, \ and\ \bibinfo {author} {\bibfnamefont {R.~A.}\ \bibnamefont {Duine}},\ }\bibfield  {title} {\enquote {\bibinfo {title} {{New perspectives for Rashba spin–orbit coupling}},}\ }\href {\doibase 10.1038/nmat4360} {\bibfield  {journal} {\bibinfo  {journal} {Nat. Mater.}\ }\textbf {\bibinfo {volume} {14}},\ \bibinfo {pages} {871--882} (\bibinfo {year} {2015})},\ \Eprint {http://arxiv.org/abs/1507.02408} {arXiv:1507.02408} \BibitemShut {NoStop}%
\bibitem [{\citenamefont {Bihlmayer}\ \emph {et~al.}(2015)\citenamefont {Bihlmayer}, \citenamefont {Rader},\ and\ \citenamefont {Winkler}}]{Bihlmayer2015}%
  \BibitemOpen
  \bibfield  {author} {\bibinfo {author} {\bibfnamefont {G.}~\bibnamefont {Bihlmayer}}, \bibinfo {author} {\bibfnamefont {O.}~\bibnamefont {Rader}}, \ and\ \bibinfo {author} {\bibfnamefont {R.}~\bibnamefont {Winkler}},\ }\bibfield  {title} {\enquote {\bibinfo {title} {{Focus on the Rashba effect}},}\ }\href {\doibase 10.1088/1367-2630/17/5/050202} {\bibfield  {journal} {\bibinfo  {journal} {New J. Phys.}\ }\textbf {\bibinfo {volume} {17}},\ \bibinfo {pages} {050202} (\bibinfo {year} {2015})}\BibitemShut {NoStop}%
\bibitem [{\citenamefont {Manchon}\ and\ \citenamefont {Zhang}(2009)}]{Manchon2009}%
  \BibitemOpen
  \bibfield  {author} {\bibinfo {author} {\bibfnamefont {A.}~\bibnamefont {Manchon}}\ and\ \bibinfo {author} {\bibfnamefont {S.}~\bibnamefont {Zhang}},\ }\bibfield  {title} {\enquote {\bibinfo {title} {{Theory of spin torque due to spin-orbit coupling}},}\ }\href {\doibase 10.1103/PHYSREVB.79.094422/FIGURES/1/THUMBNAIL} {\bibfield  {journal} {\bibinfo  {journal} {Phys. Rev. B - Condens. Matter Mater. Phys.}\ }\textbf {\bibinfo {volume} {79}},\ \bibinfo {pages} {094422} (\bibinfo {year} {2009})}\BibitemShut {NoStop}%
\bibitem [{\citenamefont {Manchon}\ \emph {et~al.}(2019)\citenamefont {Manchon}, \citenamefont {{\v{Z}}elezn{\'{y}}}, \citenamefont {Miron}, \citenamefont {Jungwirth}, \citenamefont {Sinova}, \citenamefont {Thiaville}, \citenamefont {Garello},\ and\ \citenamefont {Gambardella}}]{Manchon2019}%
  \BibitemOpen
  \bibfield  {author} {\bibinfo {author} {\bibfnamefont {A.}~\bibnamefont {Manchon}}, \bibinfo {author} {\bibfnamefont {J.}~\bibnamefont {{\v{Z}}elezn{\'{y}}}}, \bibinfo {author} {\bibfnamefont {I.~M.}\ \bibnamefont {Miron}}, \bibinfo {author} {\bibfnamefont {T.}~\bibnamefont {Jungwirth}}, \bibinfo {author} {\bibfnamefont {J.}~\bibnamefont {Sinova}}, \bibinfo {author} {\bibfnamefont {A.}~\bibnamefont {Thiaville}}, \bibinfo {author} {\bibfnamefont {K.}~\bibnamefont {Garello}}, \ and\ \bibinfo {author} {\bibfnamefont {P.}~\bibnamefont {Gambardella}},\ }\bibfield  {title} {\enquote {\bibinfo {title} {{Current-induced spin-orbit torques in ferromagnetic and antiferromagnetic systems}},}\ }\href {\doibase 10.1103/RevModPhys.91.035004} {\bibfield  {journal} {\bibinfo  {journal} {Rev. Mod. Phys.}\ }\textbf {\bibinfo {volume} {91}},\ \bibinfo {pages} {035004} (\bibinfo {year} {2019})},\ \Eprint {http://arxiv.org/abs/1801.09636} {arXiv:1801.09636} \BibitemShut {NoStop}%
\bibitem [{\citenamefont {Park}\ \emph {et~al.}(2011)\citenamefont {Park}, \citenamefont {Kim}, \citenamefont {Yu}, \citenamefont {Han},\ and\ \citenamefont {Kim}}]{Park2011}%
  \BibitemOpen
  \bibfield  {author} {\bibinfo {author} {\bibfnamefont {Seung~Ryong}\ \bibnamefont {Park}}, \bibinfo {author} {\bibfnamefont {Choong~H.}\ \bibnamefont {Kim}}, \bibinfo {author} {\bibfnamefont {Jaejun}\ \bibnamefont {Yu}}, \bibinfo {author} {\bibfnamefont {Jung~Hoon}\ \bibnamefont {Han}}, \ and\ \bibinfo {author} {\bibfnamefont {Changyoung}\ \bibnamefont {Kim}},\ }\bibfield  {title} {\enquote {\bibinfo {title} {{Orbital-angular-momentum based origin of rashba-type surface band splitting}},}\ }\href {\doibase 10.1103/PHYSREVLETT.107.156803/FIGURES/4/MEDIUM} {\bibfield  {journal} {\bibinfo  {journal} {Phys. Rev. Lett.}\ }\textbf {\bibinfo {volume} {107}},\ \bibinfo {pages} {156803} (\bibinfo {year} {2011})},\ \Eprint {http://arxiv.org/abs/1107.1554} {arXiv:1107.1554} \BibitemShut {NoStop}%
\bibitem [{\citenamefont {Park}\ \emph {et~al.}(2012)\citenamefont {Park}, \citenamefont {Kim}, \citenamefont {Rhim},\ and\ \citenamefont {Han}}]{Park2012}%
  \BibitemOpen
  \bibfield  {author} {\bibinfo {author} {\bibfnamefont {Jin~Hong}\ \bibnamefont {Park}}, \bibinfo {author} {\bibfnamefont {Choong~H.}\ \bibnamefont {Kim}}, \bibinfo {author} {\bibfnamefont {Jun~Won}\ \bibnamefont {Rhim}}, \ and\ \bibinfo {author} {\bibfnamefont {Jung~Hoon}\ \bibnamefont {Han}},\ }\bibfield  {title} {\enquote {\bibinfo {title} {{Orbital Rashba effect and its detection by circular dichroism angle-resolved photoemission spectroscopy}},}\ }\href {\doibase 10.1103/PHYSREVB.85.195401/FIGURES/3/MEDIUM} {\bibfield  {journal} {\bibinfo  {journal} {Phys. Rev. B}\ }\textbf {\bibinfo {volume} {85}},\ \bibinfo {pages} {195401} (\bibinfo {year} {2012})}\BibitemShut {NoStop}%
\bibitem [{\citenamefont {Park}\ \emph {et~al.}(2013)\citenamefont {Park}, \citenamefont {Kim}, \citenamefont {Lee},\ and\ \citenamefont {Han}}]{Park2013}%
  \BibitemOpen
  \bibfield  {author} {\bibinfo {author} {\bibfnamefont {Jin~Hong}\ \bibnamefont {Park}}, \bibinfo {author} {\bibfnamefont {Choong~H.}\ \bibnamefont {Kim}}, \bibinfo {author} {\bibfnamefont {Hyun~Woo}\ \bibnamefont {Lee}}, \ and\ \bibinfo {author} {\bibfnamefont {Jung~Hoon}\ \bibnamefont {Han}},\ }\bibfield  {title} {\enquote {\bibinfo {title} {{Orbital chirality and Rashba interaction in magnetic bands}},}\ }\href {\doibase 10.1103/PHYSREVB.87.041301/FIGURES/2/MEDIUM} {\bibfield  {journal} {\bibinfo  {journal} {Phys. Rev. B}\ }\textbf {\bibinfo {volume} {87}},\ \bibinfo {pages} {041301} (\bibinfo {year} {2013})}\BibitemShut {NoStop}%
\bibitem [{\citenamefont {Go}\ \emph {et~al.}(2017)\citenamefont {Go}, \citenamefont {Hanke}, \citenamefont {Buhl}, \citenamefont {Freimuth}, \citenamefont {Bihlmayer}, \citenamefont {Lee}, \citenamefont {Mokrousov},\ and\ \citenamefont {Bl{\"{u}}gel}}]{Go2017}%
  \BibitemOpen
  \bibfield  {author} {\bibinfo {author} {\bibfnamefont {Dongwook}\ \bibnamefont {Go}}, \bibinfo {author} {\bibfnamefont {Jan~Philipp}\ \bibnamefont {Hanke}}, \bibinfo {author} {\bibfnamefont {Patrick~M.}\ \bibnamefont {Buhl}}, \bibinfo {author} {\bibfnamefont {Frank}\ \bibnamefont {Freimuth}}, \bibinfo {author} {\bibfnamefont {Gustav}\ \bibnamefont {Bihlmayer}}, \bibinfo {author} {\bibfnamefont {Hyun~Woo}\ \bibnamefont {Lee}}, \bibinfo {author} {\bibfnamefont {Yuriy}\ \bibnamefont {Mokrousov}}, \ and\ \bibinfo {author} {\bibfnamefont {Stefan}\ \bibnamefont {Bl{\"{u}}gel}},\ }\bibfield  {title} {\enquote {\bibinfo {title} {{Toward surface orbitronics: giant orbital magnetism from the orbital Rashba effect at the surface of sp-metals}},}\ }\href {\doibase 10.1038/srep46742} {\bibfield  {journal} {\bibinfo  {journal} {Sci. Reports 2017 71}\ }\textbf {\bibinfo {volume} {7}},\ \bibinfo {pages} {46742} (\bibinfo {year} {2017})}\BibitemShut {NoStop}%
\bibitem [{\citenamefont {Go}\ \emph {et~al.}(2021{\natexlab{a}})\citenamefont {Go}, \citenamefont {Jo}, \citenamefont {Gao}, \citenamefont {Ando}, \citenamefont {Bl{\"{u}}gel}, \citenamefont {Lee},\ and\ \citenamefont {Mokrousov}}]{Go2021}%
  \BibitemOpen
  \bibfield  {author} {\bibinfo {author} {\bibfnamefont {Dongwook}\ \bibnamefont {Go}}, \bibinfo {author} {\bibfnamefont {Daegeun}\ \bibnamefont {Jo}}, \bibinfo {author} {\bibfnamefont {Tenghua}\ \bibnamefont {Gao}}, \bibinfo {author} {\bibfnamefont {Kazuya}\ \bibnamefont {Ando}}, \bibinfo {author} {\bibfnamefont {Stefan}\ \bibnamefont {Bl{\"{u}}gel}}, \bibinfo {author} {\bibfnamefont {Hyun~Woo}\ \bibnamefont {Lee}}, \ and\ \bibinfo {author} {\bibfnamefont {Yuriy}\ \bibnamefont {Mokrousov}},\ }\bibfield  {title} {\enquote {\bibinfo {title} {{Orbital Rashba effect in a surface-oxidized Cu film}},}\ }\href {\doibase 10.1103/PhysRevB.103.L121113} {\bibfield  {journal} {\bibinfo  {journal} {Phys. Rev. B}\ }\textbf {\bibinfo {volume} {103}},\ \bibinfo {pages} {L121113} (\bibinfo {year} {2021}{\natexlab{a}})}\BibitemShut {NoStop}%
\bibitem [{\citenamefont {Go}\ \emph {et~al.}(2021{\natexlab{b}})\citenamefont {Go}, \citenamefont {Jo}, \citenamefont {Lee}, \citenamefont {Kl{\"{a}}ui},\ and\ \citenamefont {Mokrousov}}]{Go2021c}%
  \BibitemOpen
  \bibfield  {author} {\bibinfo {author} {\bibfnamefont {Dongwook}\ \bibnamefont {Go}}, \bibinfo {author} {\bibfnamefont {Daegeun}\ \bibnamefont {Jo}}, \bibinfo {author} {\bibfnamefont {Hyun~Woo}\ \bibnamefont {Lee}}, \bibinfo {author} {\bibfnamefont {Mathias}\ \bibnamefont {Kl{\"{a}}ui}}, \ and\ \bibinfo {author} {\bibfnamefont {Yuriy}\ \bibnamefont {Mokrousov}},\ }\bibfield  {title} {\enquote {\bibinfo {title} {{Orbitronics: Orbital currents in solids}},}\ }\href {\doibase 10.1209/0295-5075/AC2653} {\bibfield  {journal} {\bibinfo  {journal} {Europhys. Lett.}\ }\textbf {\bibinfo {volume} {135}},\ \bibinfo {pages} {37001} (\bibinfo {year} {2021}{\natexlab{b}})},\ \Eprint {http://arxiv.org/abs/2107.08478} {arXiv:2107.08478} \BibitemShut {NoStop}%
\bibitem [{\citenamefont {Ding}\ \emph {et~al.}(2020)\citenamefont {Ding}, \citenamefont {Ross}, \citenamefont {Go}, \citenamefont {Baldrati}, \citenamefont {Ren}, \citenamefont {Freimuth}, \citenamefont {Becker}, \citenamefont {Kammerbauer}, \citenamefont {Yang}, \citenamefont {Jakob}, \citenamefont {Mokrousov},\ and\ \citenamefont {Kl{\"{a}}ui}}]{Ding2020a}%
  \BibitemOpen
  \bibfield  {author} {\bibinfo {author} {\bibfnamefont {Shilei}\ \bibnamefont {Ding}}, \bibinfo {author} {\bibfnamefont {Andrew}\ \bibnamefont {Ross}}, \bibinfo {author} {\bibfnamefont {Dongwook}\ \bibnamefont {Go}}, \bibinfo {author} {\bibfnamefont {Lorenzo}\ \bibnamefont {Baldrati}}, \bibinfo {author} {\bibfnamefont {Zengyao}\ \bibnamefont {Ren}}, \bibinfo {author} {\bibfnamefont {Frank}\ \bibnamefont {Freimuth}}, \bibinfo {author} {\bibfnamefont {Sven}\ \bibnamefont {Becker}}, \bibinfo {author} {\bibfnamefont {Fabian}\ \bibnamefont {Kammerbauer}}, \bibinfo {author} {\bibfnamefont {Jinbo}\ \bibnamefont {Yang}}, \bibinfo {author} {\bibfnamefont {Gerhard}\ \bibnamefont {Jakob}}, \bibinfo {author} {\bibfnamefont {Yuriy}\ \bibnamefont {Mokrousov}}, \ and\ \bibinfo {author} {\bibfnamefont {Mathias}\ \bibnamefont {Kl{\"{a}}ui}},\ }\bibfield  {title} {\enquote {\bibinfo {title} {{Harnessing Orbital-to-Spin Conversion of Interfacial Orbital Currents for Efficient Spin-Orbit Torques}},}\ }\href {\doibase 10.1103/PHYSREVLETT.125.177201/FIGURES/3/MEDIUM} {\bibfield  {journal} {\bibinfo  {journal} {Phys. Rev. Lett.}\ }\textbf {\bibinfo {volume} {125}},\ \bibinfo {pages} {177201} (\bibinfo {year} {2020})},\ \Eprint {http://arxiv.org/abs/2006.03649} {arXiv:2006.03649} \BibitemShut {NoStop}%
\bibitem [{\citenamefont {Ding}\ \emph {et~al.}(2022)\citenamefont {Ding}, \citenamefont {Liang}, \citenamefont {Go}, \citenamefont {Yun}, \citenamefont {Xue}, \citenamefont {Liu}, \citenamefont {Becker}, \citenamefont {Yang}, \citenamefont {Du}, \citenamefont {Wang}, \citenamefont {Yang}, \citenamefont {Jakob}, \citenamefont {Kl{\"{a}}ui}, \citenamefont {Mokrousov},\ and\ \citenamefont {Yang}}]{Ding2022a}%
  \BibitemOpen
  \bibfield  {author} {\bibinfo {author} {\bibfnamefont {Shilei}\ \bibnamefont {Ding}}, \bibinfo {author} {\bibfnamefont {Zhongyu}\ \bibnamefont {Liang}}, \bibinfo {author} {\bibfnamefont {Dongwook}\ \bibnamefont {Go}}, \bibinfo {author} {\bibfnamefont {Chao}\ \bibnamefont {Yun}}, \bibinfo {author} {\bibfnamefont {Mingzhu}\ \bibnamefont {Xue}}, \bibinfo {author} {\bibfnamefont {Zhou}\ \bibnamefont {Liu}}, \bibinfo {author} {\bibfnamefont {Sven}\ \bibnamefont {Becker}}, \bibinfo {author} {\bibfnamefont {Wenyun}\ \bibnamefont {Yang}}, \bibinfo {author} {\bibfnamefont {Honglin}\ \bibnamefont {Du}}, \bibinfo {author} {\bibfnamefont {Changsheng}\ \bibnamefont {Wang}}, \bibinfo {author} {\bibfnamefont {Yingchang}\ \bibnamefont {Yang}}, \bibinfo {author} {\bibfnamefont {Gerhard}\ \bibnamefont {Jakob}}, \bibinfo {author} {\bibfnamefont {Mathias}\ \bibnamefont {Kl{\"{a}}ui}}, \bibinfo {author} {\bibfnamefont {Yuriy}\ \bibnamefont {Mokrousov}}, \ and\ \bibinfo {author} {\bibfnamefont {Jinbo}\ \bibnamefont {Yang}},\ }\bibfield  {title} {\enquote {\bibinfo {title} {{Observation of the Orbital Rashba-Edelstein Magnetoresistance}},}\ }\href {\doibase 10.1103/PhysRevLett.128.067201} {\bibfield  {journal} {\bibinfo  {journal} {Phys. Rev. Lett.}\ }\textbf {\bibinfo {volume} {128}} (\bibinfo {year} {2022}),\ 10.1103/PhysRevLett.128.067201}\BibitemShut {NoStop}%
\bibitem [{\citenamefont {Chirolli}\ \emph {et~al.}(2022)\citenamefont {Chirolli}, \citenamefont {Mercaldo}, \citenamefont {Guarcello}, \citenamefont {Giazotto},\ and\ \citenamefont {Cuoco}}]{Chirolli2021}%
  \BibitemOpen
  \bibfield  {author} {\bibinfo {author} {\bibfnamefont {Luca}\ \bibnamefont {Chirolli}}, \bibinfo {author} {\bibfnamefont {Maria~Teresa}\ \bibnamefont {Mercaldo}}, \bibinfo {author} {\bibfnamefont {Claudio}\ \bibnamefont {Guarcello}}, \bibinfo {author} {\bibfnamefont {Francesco}\ \bibnamefont {Giazotto}}, \ and\ \bibinfo {author} {\bibfnamefont {Mario}\ \bibnamefont {Cuoco}},\ }\bibfield  {title} {\enquote {\bibinfo {title} {{Colossal orbital-Edelstein effect in non-centrosymmetric superconductors}},}\ }\href {\doibase 10.1103/PHYSREVLETT.128.217703/FIGURES/5/MEDIUM} {\bibfield  {journal} {\bibinfo  {journal} {Phys. Rev. Lett.}\ }\textbf {\bibinfo {volume} {128}},\ \bibinfo {pages} {217703} (\bibinfo {year} {2022})},\ \Eprint {http://arxiv.org/abs/2107.07476} {arXiv:2107.07476} \BibitemShut {NoStop}%
\bibitem [{\citenamefont {Ando}\ \emph {et~al.}(2024)\citenamefont {Ando}, \citenamefont {Tanaka}, \citenamefont {Cuoco}, \citenamefont {Chirolli},\ and\ \citenamefont {Mercaldo}}]{Ando2024}%
  \BibitemOpen
  \bibfield  {author} {\bibinfo {author} {\bibfnamefont {Satoshi}\ \bibnamefont {Ando}}, \bibinfo {author} {\bibfnamefont {Yukio}\ \bibnamefont {Tanaka}}, \bibinfo {author} {\bibfnamefont {Mario}\ \bibnamefont {Cuoco}}, \bibinfo {author} {\bibfnamefont {Luca}\ \bibnamefont {Chirolli}}, \ and\ \bibinfo {author} {\bibfnamefont {Maria~Teresa}\ \bibnamefont {Mercaldo}},\ }\bibfield  {title} {\enquote {\bibinfo {title} {{Spin and orbital Edelstein effect in spin-orbit coupled noncentrosymmetric superconductor}},}\ }\href {\doibase 10.1103/PHYSREVB.110.184503/FIGURES/7/MEDIUM} {\bibfield  {journal} {\bibinfo  {journal} {Phys. Rev. B}\ }\textbf {\bibinfo {volume} {110}},\ \bibinfo {pages} {184503} (\bibinfo {year} {2024})},\ \Eprint {http://arxiv.org/abs/2408.08151} {arXiv:2408.08151} \BibitemShut {NoStop}%
\bibitem [{\citenamefont {Jeon}\ \emph {et~al.}(2020)\citenamefont {Jeon}, \citenamefont {Jeon}, \citenamefont {Zhou}, \citenamefont {Migliorini}, \citenamefont {Yoon},\ and\ \citenamefont {Parkin}}]{Jeon2020}%
  \BibitemOpen
  \bibfield  {author} {\bibinfo {author} {\bibfnamefont {Kun~Rok}\ \bibnamefont {Jeon}}, \bibinfo {author} {\bibfnamefont {Jae~Chun}\ \bibnamefont {Jeon}}, \bibinfo {author} {\bibfnamefont {Xilin}\ \bibnamefont {Zhou}}, \bibinfo {author} {\bibfnamefont {Andrea}\ \bibnamefont {Migliorini}}, \bibinfo {author} {\bibfnamefont {Jiho}\ \bibnamefont {Yoon}}, \ and\ \bibinfo {author} {\bibfnamefont {Stuart~S.P.}\ \bibnamefont {Parkin}},\ }\bibfield  {title} {\enquote {\bibinfo {title} {{Giant transition-state quasiparticle spin-Hall effect in an exchange-spin-split superconductor detected by nonlocal magnon spin transport}},}\ }\href {\doibase 10.1021/ACSNANO.0C07187/SUPPL_FILE/NN0C07187_SI_001.PDF} {\bibfield  {journal} {\bibinfo  {journal} {ACS Nano}\ }\textbf {\bibinfo {volume} {14}},\ \bibinfo {pages} {15874--15883} (\bibinfo {year} {2020})}\BibitemShut {NoStop}%
\bibitem [{\citenamefont {Robbins}\ \emph {et~al.}(2020)\citenamefont {Robbins}, \citenamefont {Annett},\ and\ \citenamefont {Gradhand}}]{Robbins2020}%
  \BibitemOpen
  \bibfield  {author} {\bibinfo {author} {\bibfnamefont {J.}~\bibnamefont {Robbins}}, \bibinfo {author} {\bibfnamefont {J.~F.}\ \bibnamefont {Annett}}, \ and\ \bibinfo {author} {\bibfnamefont {M.}~\bibnamefont {Gradhand}},\ }\bibfield  {title} {\enquote {\bibinfo {title} {{Theory of the orbital moment in a superconductor}},}\ }\href {\doibase 10.1103/PhysRevB.101.134505} {\bibfield  {journal} {\bibinfo  {journal} {Phys. Rev. B}\ }\textbf {\bibinfo {volume} {101}},\ \bibinfo {pages} {134505} (\bibinfo {year} {2020})}\BibitemShut {NoStop}%
\bibitem [{\citenamefont {Saunderson}\ \emph {et~al.}(2020{\natexlab{a}})\citenamefont {Saunderson}, \citenamefont {Annett}, \citenamefont {{\'{U}}jfalussy}, \citenamefont {Csire},\ and\ \citenamefont {Gradhand}}]{Saunderson2020}%
  \BibitemOpen
  \bibfield  {author} {\bibinfo {author} {\bibfnamefont {Tom~G.}\ \bibnamefont {Saunderson}}, \bibinfo {author} {\bibfnamefont {James~F.}\ \bibnamefont {Annett}}, \bibinfo {author} {\bibfnamefont {Bal{\'{a}}zs}\ \bibnamefont {{\'{U}}jfalussy}}, \bibinfo {author} {\bibfnamefont {G{\'{a}}bor}\ \bibnamefont {Csire}}, \ and\ \bibinfo {author} {\bibfnamefont {Martin}\ \bibnamefont {Gradhand}},\ }\bibfield  {title} {\enquote {\bibinfo {title} {{Gap Anisotropy in Multiband Superconductors Based on Multiple Scattering Theory}},}\ }\href@noop {} {\bibfield  {journal} {\bibinfo  {journal} {Phys. Rev. B}\ }\textbf {\bibinfo {volume} {101}},\ \bibinfo {pages} {064510} (\bibinfo {year} {2020}{\natexlab{a}})}\BibitemShut {NoStop}%
\bibitem [{\citenamefont {Saunderson}\ \emph {et~al.}(2020{\natexlab{b}})\citenamefont {Saunderson}, \citenamefont {Győrgyp{\'{a}}l}, \citenamefont {Annett}, \citenamefont {Csire}, \citenamefont {{\'{U}}jfalussy},\ and\ \citenamefont {Gradhand}}]{Saunderson2020b}%
  \BibitemOpen
  \bibfield  {author} {\bibinfo {author} {\bibfnamefont {Tom~G.}\ \bibnamefont {Saunderson}}, \bibinfo {author} {\bibfnamefont {Zsolt}\ \bibnamefont {Győrgyp{\'{a}}l}}, \bibinfo {author} {\bibfnamefont {James~F.}\ \bibnamefont {Annett}}, \bibinfo {author} {\bibfnamefont {G{\'{a}}bor}\ \bibnamefont {Csire}}, \bibinfo {author} {\bibfnamefont {Bal{\'{a}}zs}\ \bibnamefont {{\'{U}}jfalussy}}, \ and\ \bibinfo {author} {\bibfnamefont {Martin}\ \bibnamefont {Gradhand}},\ }\bibfield  {title} {\enquote {\bibinfo {title} {{Real-space multiple scattering theory for superconductors with impurities}},}\ }\href {\doibase 10.1103/PhysRevB.102.245106} {\bibfield  {journal} {\bibinfo  {journal} {Phys. Rev. B}\ }\textbf {\bibinfo {volume} {102}},\ \bibinfo {pages} {245106} (\bibinfo {year} {2020}{\natexlab{b}})},\ \Eprint {http://arxiv.org/abs/2009.08766} {arXiv:2009.08766} \BibitemShut {NoStop}%
\bibitem [{\citenamefont {Saunderson}\ \emph {et~al.}(2022)\citenamefont {Saunderson}, \citenamefont {Annett}, \citenamefont {Csire},\ and\ \citenamefont {Gradhand}}]{Saunderson2022}%
  \BibitemOpen
  \bibfield  {author} {\bibinfo {author} {\bibfnamefont {Tom~G.}\ \bibnamefont {Saunderson}}, \bibinfo {author} {\bibfnamefont {James~F.}\ \bibnamefont {Annett}}, \bibinfo {author} {\bibfnamefont {G{\'{a}}bor}\ \bibnamefont {Csire}}, \ and\ \bibinfo {author} {\bibfnamefont {Martin}\ \bibnamefont {Gradhand}},\ }\bibfield  {title} {\enquote {\bibinfo {title} {{Full orbital decomposition of Yu-Shiba-Rusinov states based on first principles}},}\ }\href {\doibase 10.1103/PhysRevB.105.014424} {\bibfield  {journal} {\bibinfo  {journal} {Phys. Rev. B}\ }\textbf {\bibinfo {volume} {105}},\ \bibinfo {pages} {014424} (\bibinfo {year} {2022})}\BibitemShut {NoStop}%
\bibitem [{\citenamefont {Wu}\ \emph {et~al.}(2023)\citenamefont {Wu}, \citenamefont {Thill}, \citenamefont {Crosbie}, \citenamefont {Saunderson},\ and\ \citenamefont {Gradhand}}]{Wu2023}%
  \BibitemOpen
  \bibfield  {author} {\bibinfo {author} {\bibfnamefont {Ming-Hung}\ \bibnamefont {Wu}}, \bibinfo {author} {\bibfnamefont {Emma}\ \bibnamefont {Thill}}, \bibinfo {author} {\bibfnamefont {Jacob}\ \bibnamefont {Crosbie}}, \bibinfo {author} {\bibfnamefont {Tom~G.}\ \bibnamefont {Saunderson}}, \ and\ \bibinfo {author} {\bibfnamefont {Martin}\ \bibnamefont {Gradhand}},\ }\bibfield  {title} {\enquote {\bibinfo {title} {{Magnetic impurities on superconducting Pb surfaces}},}\ }\href {\doibase 10.1103/PhysRevB.107.094409} {\bibfield  {journal} {\bibinfo  {journal} {Phys. Rev. B}\ }\textbf {\bibinfo {volume} {107}},\ \bibinfo {pages} {094409} (\bibinfo {year} {2023})}\BibitemShut {NoStop}%
\bibitem [{\citenamefont {Csire}\ \emph {et~al.}(2018{\natexlab{a}})\citenamefont {Csire}, \citenamefont {De{\'{a}}k}, \citenamefont {Ny{\'{a}}ri}, \citenamefont {Ebert}, \citenamefont {Annett},\ and\ \citenamefont {{\'{U}}jfalussy}}]{Csire2018}%
  \BibitemOpen
  \bibfield  {author} {\bibinfo {author} {\bibfnamefont {G.}~\bibnamefont {Csire}}, \bibinfo {author} {\bibfnamefont {A.}~\bibnamefont {De{\'{a}}k}}, \bibinfo {author} {\bibfnamefont {B.}~\bibnamefont {Ny{\'{a}}ri}}, \bibinfo {author} {\bibfnamefont {H.}~\bibnamefont {Ebert}}, \bibinfo {author} {\bibfnamefont {J.~F.}\ \bibnamefont {Annett}}, \ and\ \bibinfo {author} {\bibfnamefont {B.}~\bibnamefont {{\'{U}}jfalussy}},\ }\bibfield  {title} {\enquote {\bibinfo {title} {{Relativistic spin-polarized KKR theory for superconducting heterostructures: Oscillating order parameter in the Au layer of Nb/Au/Fe trilayers}},}\ }\href {\doibase 10.1103/PhysRevB.97.024514} {\bibfield  {journal} {\bibinfo  {journal} {Phys. Rev. B}\ }\textbf {\bibinfo {volume} {97}},\ \bibinfo {pages} {024514} (\bibinfo {year} {2018}{\natexlab{a}})}\BibitemShut {NoStop}%
\bibitem [{\citenamefont {Sato}\ and\ \citenamefont {Ando}(2017)}]{Sato2017}%
  \BibitemOpen
  \bibfield  {author} {\bibinfo {author} {\bibfnamefont {M.}~\bibnamefont {Sato}}\ and\ \bibinfo {author} {\bibfnamefont {Y.}~\bibnamefont {Ando}},\ }\bibfield  {title} {\enquote {\bibinfo {title} {{Topological superconductors: A review}},}\ }\href {\doibase 10.1088/1361-6633/aa6ac7} {\bibfield  {journal} {\bibinfo  {journal} {Reports Prog. Phys.}\ }\textbf {\bibinfo {volume} {80}},\ \bibinfo {pages} {076501} (\bibinfo {year} {2017})},\ \Eprint {http://arxiv.org/abs/1608.03395} {arXiv:1608.03395} \BibitemShut {NoStop}%
\bibitem [{\citenamefont {Csire}\ \emph {et~al.}(2018{\natexlab{b}})\citenamefont {Csire}, \citenamefont {{\'{U}}jfalussy},\ and\ \citenamefont {Annett}}]{Csire2018b}%
  \BibitemOpen
  \bibfield  {author} {\bibinfo {author} {\bibfnamefont {G{\'{a}}bor}\ \bibnamefont {Csire}}, \bibinfo {author} {\bibfnamefont {Bal{\'{a}}zs}\ \bibnamefont {{\'{U}}jfalussy}}, \ and\ \bibinfo {author} {\bibfnamefont {James~F.}\ \bibnamefont {Annett}},\ }\bibfield  {title} {\enquote {\bibinfo {title} {{Nonunitary triplet pairing in the noncentrosymmetric superconductor LaNiC2}},}\ }\href {\doibase 10.1140/EPJB/E2018-90095-7} {\bibfield  {journal} {\bibinfo  {journal} {Eur. Phys. J. B}\ }\textbf {\bibinfo {volume} {91}} (\bibinfo {year} {2018}{\natexlab{b}}),\ 10.1140/EPJB/E2018-90095-7}\BibitemShut {NoStop}%
\bibitem [{\citenamefont {Csire}\ \emph {et~al.}(2022)\citenamefont {Csire}, \citenamefont {Annett}, \citenamefont {Quintanilla},\ and\ \citenamefont {{\'{U}}jfalussy}}]{Csire2020a}%
  \BibitemOpen
  \bibfield  {author} {\bibinfo {author} {\bibfnamefont {G{\'{a}}bor}\ \bibnamefont {Csire}}, \bibinfo {author} {\bibfnamefont {James~F}\ \bibnamefont {Annett}}, \bibinfo {author} {\bibfnamefont {Jorge}\ \bibnamefont {Quintanilla}}, \ and\ \bibinfo {author} {\bibfnamefont {Bal{\'{a}}zs}\ \bibnamefont {{\'{U}}jfalussy}},\ }\bibfield  {title} {\enquote {\bibinfo {title} {{Magnetically-textured superconductivity in elemental Rhenium}},}\ }\href {\doibase 10.48550/arxiv.2005.05702} {\bibfield  {journal} {\bibinfo  {journal} {Phys. Rev. B}\ }\textbf {\bibinfo {volume} {106}},\ \bibinfo {pages} {L020501} (\bibinfo {year} {2022})},\ \Eprint {http://arxiv.org/abs/2005.05702} {arXiv:2005.05702} \BibitemShut {NoStop}%
\bibitem [{\citenamefont {Ghosh}\ \emph {et~al.}(2020)\citenamefont {Ghosh}, \citenamefont {Csire}, \citenamefont {Whittlesea}, \citenamefont {Annett}, \citenamefont {Gradhand}, \citenamefont {{\'{U}}jfalussy},\ and\ \citenamefont {Quintanilla}}]{Ghosh2020b}%
  \BibitemOpen
  \bibfield  {author} {\bibinfo {author} {\bibfnamefont {Sudeep~Kumar}\ \bibnamefont {Ghosh}}, \bibinfo {author} {\bibfnamefont {G{\'{a}}bor}\ \bibnamefont {Csire}}, \bibinfo {author} {\bibfnamefont {Philip}\ \bibnamefont {Whittlesea}}, \bibinfo {author} {\bibfnamefont {James~F.}\ \bibnamefont {Annett}}, \bibinfo {author} {\bibfnamefont {Martin}\ \bibnamefont {Gradhand}}, \bibinfo {author} {\bibfnamefont {Bal{\'{a}}zs}\ \bibnamefont {{\'{U}}jfalussy}}, \ and\ \bibinfo {author} {\bibfnamefont {Jorge}\ \bibnamefont {Quintanilla}},\ }\bibfield  {title} {\enquote {\bibinfo {title} {{Quantitative theory of triplet pairing in the unconventional superconductor LaNiGa2}},}\ }\href {\doibase 10.1103/PHYSREVB.101.100506/FIGURES/3/MEDIUM} {\bibfield  {journal} {\bibinfo  {journal} {Phys. Rev. B}\ }\textbf {\bibinfo {volume} {101}},\ \bibinfo {pages} {100506} (\bibinfo {year} {2020})},\ \Eprint {http://arxiv.org/abs/1912.08160} {arXiv:1912.08160} \BibitemShut {NoStop}%
\bibitem [{\citenamefont {R{\"{u}}{\ss}mann}\ and\ \citenamefont {Bl{\"{u}}gel}(2022{\natexlab{a}})}]{Rußmann2022b}%
  \BibitemOpen
  \bibfield  {author} {\bibinfo {author} {\bibfnamefont {Philipp}\ \bibnamefont {R{\"{u}}{\ss}mann}}\ and\ \bibinfo {author} {\bibfnamefont {Stefan}\ \bibnamefont {Bl{\"{u}}gel}},\ }\bibfield  {title} {\enquote {\bibinfo {title} {{Density functional Bogoliubov-de Gennes analysis of superconducting Nb and Nb(110) surfaces}},}\ }\href {\doibase 10.1103/PHYSREVB.105.125143/FIGURES/6/MEDIUM} {\bibfield  {journal} {\bibinfo  {journal} {Phys. Rev. B}\ }\textbf {\bibinfo {volume} {105}},\ \bibinfo {pages} {125143} (\bibinfo {year} {2022}{\natexlab{a}})},\ \Eprint {http://arxiv.org/abs/2110.01713} {arXiv:2110.01713} \BibitemShut {NoStop}%
\bibitem [{\citenamefont {R{\"{u}}{\ss}mann}\ \emph {et~al.}(2023)\citenamefont {R{\"{u}}{\ss}mann}, \citenamefont {Bahari}, \citenamefont {Bl{\"{u}}gel},\ and\ \citenamefont {Trauzettel}}]{Rußmann2023}%
  \BibitemOpen
  \bibfield  {author} {\bibinfo {author} {\bibfnamefont {Philipp}\ \bibnamefont {R{\"{u}}{\ss}mann}}, \bibinfo {author} {\bibfnamefont {Masoud}\ \bibnamefont {Bahari}}, \bibinfo {author} {\bibfnamefont {Stefan}\ \bibnamefont {Bl{\"{u}}gel}}, \ and\ \bibinfo {author} {\bibfnamefont {Bj{\"{o}}rn}\ \bibnamefont {Trauzettel}},\ }\bibfield  {title} {\enquote {\bibinfo {title} {{Interorbital Cooper pairing at finite energies in Rashba surface states}},}\ }\href {\doibase 10.1103/PHYSREVRESEARCH.5.043181/FIGURES/9/MEDIUM} {\bibfield  {journal} {\bibinfo  {journal} {Phys. Rev. Res.}\ }\textbf {\bibinfo {volume} {5}},\ \bibinfo {pages} {043181} (\bibinfo {year} {2023})},\ \Eprint {http://arxiv.org/abs/2307.13990} {arXiv:2307.13990} \BibitemShut {NoStop}%
\bibitem [{\citenamefont {Yamazaki}\ \emph {et~al.}(2024)\citenamefont {Yamazaki}, \citenamefont {Csire}, \citenamefont {Kucska}, \citenamefont {Shannon}, \citenamefont {Takagi},\ and\ \citenamefont {Bal{\'{a}}zs´ujfalussy}}]{Yamazaki2024}%
  \BibitemOpen
  \bibfield  {author} {\bibinfo {author} {\bibfnamefont {Hiroki}\ \bibnamefont {Yamazaki}}, \bibinfo {author} {\bibfnamefont {G{\'{a}}bor}\ \bibnamefont {Csire}}, \bibinfo {author} {\bibfnamefont {N{\'{o}}ra}\ \bibnamefont {Kucska}}, \bibinfo {author} {\bibfnamefont {Nic}\ \bibnamefont {Shannon}}, \bibinfo {author} {\bibfnamefont {Hidenori}\ \bibnamefont {Takagi}}, \ and\ \bibinfo {author} {\bibfnamefont {Bal{\'{a}}zs~Bal{\'{a}}zs´}\ \bibnamefont {Bal{\'{a}}zs´ujfalussy}},\ }\bibfield  {title} {\enquote {\bibinfo {title} {{Quantum size effects on Andreev transport in Nb/Au/Nb Josephson junctions: A combined ab-initio and experimental study}},}\ }\href {https://arxiv.org/abs/2404.09784v1} {\  (\bibinfo {year} {2024})},\ \Eprint {http://arxiv.org/abs/2404.09784} {arXiv:2404.09784} \BibitemShut {NoStop}%
\bibitem [{\citenamefont {Reho}\ \emph {et~al.}(2024{\natexlab{a}})\citenamefont {Reho}, \citenamefont {Wittemeier}, \citenamefont {Kole}, \citenamefont {Ordej{\'{o}}n},\ and\ \citenamefont {Zanolli}}]{Reho2024}%
  \BibitemOpen
  \bibfield  {author} {\bibinfo {author} {\bibfnamefont {R.}~\bibnamefont {Reho}}, \bibinfo {author} {\bibfnamefont {N.}~\bibnamefont {Wittemeier}}, \bibinfo {author} {\bibfnamefont {A.~H.}\ \bibnamefont {Kole}}, \bibinfo {author} {\bibfnamefont {P.}~\bibnamefont {Ordej{\'{o}}n}}, \ and\ \bibinfo {author} {\bibfnamefont {Z.}~\bibnamefont {Zanolli}},\ }\bibfield  {title} {\enquote {\bibinfo {title} {{Density functional Bogoliubov-de Gennes theory for superconductors implemented in the SIESTA code}},}\ }\href {\doibase 10.1103/PHYSREVB.110.134505/FIGURES/15/MEDIUM} {\bibfield  {journal} {\bibinfo  {journal} {Phys. Rev. B}\ }\textbf {\bibinfo {volume} {110}},\ \bibinfo {pages} {134505} (\bibinfo {year} {2024}{\natexlab{a}})},\ \Eprint {http://arxiv.org/abs/2406.02022} {arXiv:2406.02022} \BibitemShut {NoStop}%
\bibitem [{\citenamefont {Ny{\'{a}}ri}\ \emph {et~al.}(2023)\citenamefont {Ny{\'{a}}ri}, \citenamefont {L{\'{a}}szl{\'{o}}ffy}, \citenamefont {Csire}, \citenamefont {Szunyogh},\ and\ \citenamefont {{\'{U}}jfalussy}}]{Nyari2023}%
  \BibitemOpen
  \bibfield  {author} {\bibinfo {author} {\bibfnamefont {Bendeg{\'{u}}z}\ \bibnamefont {Ny{\'{a}}ri}}, \bibinfo {author} {\bibfnamefont {Andr{\'{a}}s}\ \bibnamefont {L{\'{a}}szl{\'{o}}ffy}}, \bibinfo {author} {\bibfnamefont {G{\'{a}}bor}\ \bibnamefont {Csire}}, \bibinfo {author} {\bibfnamefont {L{\'{a}}szl{\'{o}}}\ \bibnamefont {Szunyogh}}, \ and\ \bibinfo {author} {\bibfnamefont {Bal{\'{a}}zs}\ \bibnamefont {{\'{U}}jfalussy}},\ }\bibfield  {title} {\enquote {\bibinfo {title} {{Topological superconductivity from first principles. I. Shiba band structure and topological edge states of artificial spin chains}},}\ }\href {\doibase 10.1103/PHYSREVB.108.134512/FIGURES/4/MEDIUM} {\bibfield  {journal} {\bibinfo  {journal} {Phys. Rev. B}\ }\textbf {\bibinfo {volume} {108}},\ \bibinfo {pages} {134512} (\bibinfo {year} {2023})},\ \Eprint {http://arxiv.org/abs/2308.13824} {arXiv:2308.13824} \BibitemShut {NoStop}%
\bibitem [{\citenamefont {L{\'{a}}szl{\'{o}}ffy}\ \emph {et~al.}(2023)\citenamefont {L{\'{a}}szl{\'{o}}ffy}, \citenamefont {Ny{\'{a}}ri}, \citenamefont {Csire}, \citenamefont {Szunyogh},\ and\ \citenamefont {{\'{U}}jfalussy}}]{Laszloffy2023}%
  \BibitemOpen
  \bibfield  {author} {\bibinfo {author} {\bibfnamefont {Andr{\'{a}}s}\ \bibnamefont {L{\'{a}}szl{\'{o}}ffy}}, \bibinfo {author} {\bibfnamefont {Bendeg{\'{u}}z}\ \bibnamefont {Ny{\'{a}}ri}}, \bibinfo {author} {\bibfnamefont {G{\'{a}}bor}\ \bibnamefont {Csire}}, \bibinfo {author} {\bibfnamefont {L{\'{a}}szl{\'{o}}}\ \bibnamefont {Szunyogh}}, \ and\ \bibinfo {author} {\bibfnamefont {Bal{\'{a}}zs}\ \bibnamefont {{\'{U}}jfalussy}},\ }\bibfield  {title} {\enquote {\bibinfo {title} {{Topological superconductivity from first principles. II. Effects from manipulation of spin spirals: Topological fragmentation, braiding, and quasi-Majorana bound states}},}\ }\href {\doibase 10.1103/PHYSREVB.108.134513/FIGURES/9/MEDIUM} {\bibfield  {journal} {\bibinfo  {journal} {Phys. Rev. B}\ }\textbf {\bibinfo {volume} {108}},\ \bibinfo {pages} {134513} (\bibinfo {year} {2023})}\BibitemShut {NoStop}%
\bibitem [{\citenamefont {R{\"{u}}{\ss}mann}\ and\ \citenamefont {Bl{\"{u}}gel}(2022{\natexlab{b}})}]{Rußmann2022a}%
  \BibitemOpen
  \bibfield  {author} {\bibinfo {author} {\bibfnamefont {Philipp}\ \bibnamefont {R{\"{u}}{\ss}mann}}\ and\ \bibinfo {author} {\bibfnamefont {Stefan}\ \bibnamefont {Bl{\"{u}}gel}},\ }\bibfield  {title} {\enquote {\bibinfo {title} {{Proximity induced superconductivity in a topological insulator}},}\ }\href {\doibase 10.48550/arxiv.2208.14289} {\  (\bibinfo {year} {2022}{\natexlab{b}}),\ 10.48550/arxiv.2208.14289},\ \Eprint {http://arxiv.org/abs/2208.14289} {arXiv:2208.14289} \BibitemShut {NoStop}%
\bibitem [{\citenamefont {Reho}\ \emph {et~al.}(2024{\natexlab{b}})\citenamefont {Reho}, \citenamefont {Botello-M{\'{e}}ndez},\ and\ \citenamefont {Zanolli}}]{Reho2024a}%
  \BibitemOpen
  \bibfield  {author} {\bibinfo {author} {\bibfnamefont {R.}~\bibnamefont {Reho}}, \bibinfo {author} {\bibfnamefont {A.~R.}\ \bibnamefont {Botello-M{\'{e}}ndez}}, \ and\ \bibinfo {author} {\bibfnamefont {Zeila}\ \bibnamefont {Zanolli}},\ }\bibfield  {title} {\enquote {\bibinfo {title} {{Ab initio study of Proximity-Induced Superconductivity in PbTe/Pb heterostructures}},}\ }\href {https://arxiv.org/abs/2412.01749v1} {\  (\bibinfo {year} {2024}{\natexlab{b}})},\ \Eprint {http://arxiv.org/abs/2412.01749} {arXiv:2412.01749} \BibitemShut {NoStop}%
\bibitem [{\citenamefont {Park}\ \emph {et~al.}(2020)\citenamefont {Park}, \citenamefont {Csire},\ and\ \citenamefont {Ujfalussy}}]{Park2020a}%
  \BibitemOpen
  \bibfield  {author} {\bibinfo {author} {\bibfnamefont {Kyungwha}\ \bibnamefont {Park}}, \bibinfo {author} {\bibfnamefont {Gabor}\ \bibnamefont {Csire}}, \ and\ \bibinfo {author} {\bibfnamefont {Balazs}\ \bibnamefont {Ujfalussy}},\ }\bibfield  {title} {\enquote {\bibinfo {title} {{Proximity effect in a superconductor-topological insulator heterostructure based on first principles}},}\ }\href {\doibase 10.1103/PHYSREVB.102.134504/FIGURES/5/MEDIUM} {\bibfield  {journal} {\bibinfo  {journal} {Phys. Rev. B}\ }\textbf {\bibinfo {volume} {102}},\ \bibinfo {pages} {134504} (\bibinfo {year} {2020})},\ \Eprint {http://arxiv.org/abs/2005.02570} {arXiv:2005.02570} \BibitemShut {NoStop}%
\bibitem [{Sup()}]{Supplementary}%
  \BibitemOpen
  \href@noop {} {}\bibinfo {note} {See Supplemental Material, which also contains Refs.~\cite{Csire2018,Capelle1999,Capelle1999a,Zabloudil2004,Gradhand2009,Ebert1996,Ebert2016,Saunderson2020,Saunderson2020b,Saunderson2022,Vosko1980,Chirolli2021}}\BibitemShut {NoStop}%
\bibitem [{\citenamefont {Capelle}\ and\ \citenamefont {Gross}(1999{\natexlab{a}})}]{Capelle1999}%
  \BibitemOpen
  \bibfield  {author} {\bibinfo {author} {\bibfnamefont {K.}~\bibnamefont {Capelle}}\ and\ \bibinfo {author} {\bibfnamefont {E.~K.~U.}\ \bibnamefont {Gross}},\ }\bibfield  {title} {\enquote {\bibinfo {title} {{Relativistic framework for microscopic theories of superconductivity. II. The Pauli equation for superconductors}},}\ }\href {\doibase 10.1103/PhysRevB.59.7155} {\bibfield  {journal} {\bibinfo  {journal} {Phys. Rev. B}\ }\textbf {\bibinfo {volume} {59}},\ \bibinfo {pages} {7155--7165} (\bibinfo {year} {1999}{\natexlab{a}})}\BibitemShut {NoStop}%
\bibitem [{\citenamefont {Capelle}\ and\ \citenamefont {Gross}(1999{\natexlab{b}})}]{Capelle1999a}%
  \BibitemOpen
  \bibfield  {author} {\bibinfo {author} {\bibfnamefont {K.}~\bibnamefont {Capelle}}\ and\ \bibinfo {author} {\bibfnamefont {E.~K.~U.}\ \bibnamefont {Gross}},\ }\bibfield  {title} {\enquote {\bibinfo {title} {{Relativistic framework for microscopic theories of superconductivity. I. The Dirac equation for superconductors}},}\ }\href {\doibase 10.1103/PhysRevB.59.7155} {\bibfield  {journal} {\bibinfo  {journal} {Phys. Rev. B}\ }\textbf {\bibinfo {volume} {59}},\ \bibinfo {pages} {7140--7154} (\bibinfo {year} {1999}{\natexlab{b}})}\BibitemShut {NoStop}%
\bibitem [{\citenamefont {Zabloudil}\ \emph {et~al.}(2004)\citenamefont {Zabloudil}, \citenamefont {Hammerling}, \citenamefont {Szunyogh},\ and\ \citenamefont {Weinberger}}]{Zabloudil2004}%
  \BibitemOpen
  \bibfield  {author} {\bibinfo {author} {\bibfnamefont {J.}~\bibnamefont {Zabloudil}}, \bibinfo {author} {\bibfnamefont {R.}~\bibnamefont {Hammerling}}, \bibinfo {author} {\bibfnamefont {L.}~\bibnamefont {Szunyogh}}, \ and\ \bibinfo {author} {\bibfnamefont {P.}~\bibnamefont {Weinberger}},\ }\href {http://books.google.com/books?hl=en&lr=&id=j8hVhaVyx7EC&oi=fnd&pg=PA1&dq=Electron+Scattering+in+Solid+Matter+A+Theoretical+and+Computational+Treatise&ots=t96tKV-LJ-&sig=U2B-RH4dkxnAuqIktz2l1940FV4} {\emph {\bibinfo {title} {Springer}}}\ (\bibinfo  {publisher} {Springer Berlin Heidelberg New York},\ \bibinfo {year} {2004})\BibitemShut {NoStop}%
\bibitem [{\citenamefont {Gradhand}\ \emph {et~al.}(2009)\citenamefont {Gradhand}, \citenamefont {Czerner}, \citenamefont {Fedorov}, \citenamefont {Zahn}, \citenamefont {Yavorsky}, \citenamefont {Szunyogh},\ and\ \citenamefont {Mertig}}]{Gradhand2009}%
  \BibitemOpen
  \bibfield  {author} {\bibinfo {author} {\bibfnamefont {Martin}\ \bibnamefont {Gradhand}}, \bibinfo {author} {\bibfnamefont {Michael}\ \bibnamefont {Czerner}}, \bibinfo {author} {\bibfnamefont {Dmitry~V.}\ \bibnamefont {Fedorov}}, \bibinfo {author} {\bibfnamefont {Peter}\ \bibnamefont {Zahn}}, \bibinfo {author} {\bibfnamefont {Bogdan~Yu}\ \bibnamefont {Yavorsky}}, \bibinfo {author} {\bibfnamefont {L{\'{a}}szlo}\ \bibnamefont {Szunyogh}}, \ and\ \bibinfo {author} {\bibfnamefont {Ingrid}\ \bibnamefont {Mertig}},\ }\bibfield  {title} {\enquote {\bibinfo {title} {{Spin polarization on Fermi surfaces of metals by the KKR method}},}\ }\href {\doibase 10.1103/PhysRevB.80.224413} {\bibfield  {journal} {\bibinfo  {journal} {Phys. Rev. B}\ }\textbf {\bibinfo {volume} {80}},\ \bibinfo {pages} {224413} (\bibinfo {year} {2009})}\BibitemShut {NoStop}%
\bibitem [{\citenamefont {Ebert}\ \emph {et~al.}(1996)\citenamefont {Ebert}, \citenamefont {Freyer}, \citenamefont {Vernes},\ and\ \citenamefont {Guo}}]{Ebert1996}%
  \BibitemOpen
  \bibfield  {author} {\bibinfo {author} {\bibfnamefont {H.}~\bibnamefont {Ebert}}, \bibinfo {author} {\bibfnamefont {H.}~\bibnamefont {Freyer}}, \bibinfo {author} {\bibfnamefont {A.}~\bibnamefont {Vernes}}, \ and\ \bibinfo {author} {\bibfnamefont {G.}~\bibnamefont {Guo}},\ }\bibfield  {title} {\enquote {\bibinfo {title} {{Manipulation of the spin-orbit coupling using the Dirac equation for spin-dependent potentials}},}\ }\href {\doibase 10.1103/PhysRevB.53.7721} {\bibfield  {journal} {\bibinfo  {journal} {Phys. Rev. B}\ }\textbf {\bibinfo {volume} {53}},\ \bibinfo {pages} {7721} (\bibinfo {year} {1996})}\BibitemShut {NoStop}%
\bibitem [{\citenamefont {Ebert}\ \emph {et~al.}(2016)\citenamefont {Ebert}, \citenamefont {Braun}, \citenamefont {K{\"{o}}dderitzsch},\ and\ \citenamefont {Mankovsky}}]{Ebert2016}%
  \BibitemOpen
  \bibfield  {author} {\bibinfo {author} {\bibfnamefont {H.}~\bibnamefont {Ebert}}, \bibinfo {author} {\bibfnamefont {J.}~\bibnamefont {Braun}}, \bibinfo {author} {\bibfnamefont {D.}~\bibnamefont {K{\"{o}}dderitzsch}}, \ and\ \bibinfo {author} {\bibfnamefont {S.}~\bibnamefont {Mankovsky}},\ }\bibfield  {title} {\enquote {\bibinfo {title} {{Fully relativistic multiple scattering calculations for general potentials}},}\ }\href {\doibase 10.1103/PHYSREVB.93.075145/FIGURES/2/MEDIUM} {\bibfield  {journal} {\bibinfo  {journal} {Phys. Rev. B}\ }\textbf {\bibinfo {volume} {93}},\ \bibinfo {pages} {075145} (\bibinfo {year} {2016})}\BibitemShut {NoStop}%
\bibitem [{\citenamefont {Vosko}\ \emph {et~al.}(1980)\citenamefont {Vosko}, \citenamefont {Wilk},\ and\ \citenamefont {Nusaur}}]{Vosko1980}%
  \BibitemOpen
  \bibfield  {author} {\bibinfo {author} {\bibfnamefont {S.~H.}\ \bibnamefont {Vosko}}, \bibinfo {author} {\bibfnamefont {L.}~\bibnamefont {Wilk}}, \ and\ \bibinfo {author} {\bibfnamefont {M.}~\bibnamefont {Nusaur}},\ }\bibfield  {title} {\enquote {\bibinfo {title} {{Accurate spin-dependent electron liquid correlation energies for local spin density calculations: a critical analysis1}},}\ }\href@noop {} {\bibfield  {journal} {\bibinfo  {journal} {Can. J. Phys.}\ }\textbf {\bibinfo {volume} {58}},\ \bibinfo {pages} {1200} (\bibinfo {year} {1980})}\BibitemShut {NoStop}%
\end{thebibliography}%


%


\end{document}